%% file: main.tex
\documentclass[%
  conference,%
  10pt,%
]{IEEEtran}

\usepackage{cite}
\usepackage{amsmath,amssymb,amsfonts}
\allowdisplaybreaks
\DeclareMathOperator{\verdict}{verdict}

\usepackage{booktabs}
\usepackage{url}

\input{defs}

\newcommand{\ourtool}{\toolname{FlaPy}}

\begin{document}

% \title{Python Flaky Tests}
% \title{Flaky Tests in Open Source Python Projects:\\ An Empirical Study}
\title{An Empirical Study of Flaky Tests in Python}

\author{%
 \IEEEauthorblockN{Martin Gruber}%
 \IEEEauthorblockA{%
   \textit{BMW Group}\\%
   Munich, Germany\\%
   martin.gr.gruber@bmw.de%
 }%
 \and%
 \IEEEauthorblockN{Stephan Lukasczyk}%
 \IEEEauthorblockA{%
   \textit{University of Passau}\\%
   Passau, Germany\\%
   stephan.lukasczyk@uni-passau.de%
 }%
 \and%
 \IEEEauthorblockN{Florian Kroi\ss}%
 \IEEEauthorblockA{%
   \textit{University of Passau}\\%
   Passau, Germany\\%
   kroiss@fim.uni-passau.de%
 }%
 \and%
 \IEEEauthorblockN{Gordon Fraser}%
 \IEEEauthorblockA{%
   \textit{University of Passau}\\%
   Passau, Germany\\%
   gordon.fraser@uni-passau.de%
 }
}

% \author{\IEEEauthorblockN{Anonymous Author(s)}}

\maketitle

% -- NUMBER MARCOS

\newcommand{\numAllProjects}{\num{22352}\xspace}
\newcommand{\numAllTests}{\num{876186}\xspace}

\newcommand{\numFlakyProjects}{\num{1006}\xspace}
\newcommand{\numNODFlakyProjects}{\num{279}\xspace}

\newcommand{\FlakyProjectsOverAllProjects}{\SI{4.5}{\percent}\xspace}
\newcommand{\FlakyTestsOverAllTests}{\SI{0.86}{\percent}\xspace}

\newcommand{\numNODFlakyTests}{\num{952}\xspace}
\newcommand{\numODFlakyTests}{\num{4461}\xspace}
\newcommand{\numInfrastructureFlakyTests}{\num{2158}\xspace}
\newcommand{\numFlakyTests}{\num{7571}\xspace}

\newcommand{\numODVictimFlakyTests}{\num{3168}\xspace}
\newcommand{\numODBrittleFlakyTests}{\num{738}\xspace}
\newcommand{\numODNotAnalysedFlakyTests}{\num{555}\xspace}

\newcommand{\NodFlakyTestsOverAllFlakyTests}{\SI{13}{\percent}\xspace}
\newcommand{\OdFlakyTestsOverAllFlakyTests}{\SI{59}{\percent}\xspace}
\newcommand{\InfrastructureFlakyTestsOverAllFlakyTests}{\SI{28}{\percent}\xspace}

% no. Tests per Project
\newcommand{\numMinTestsAllProjects}{\num{1}\xspace}
\newcommand{\numMaxTestsAllProjects}{\num{76301}\xspace}
\newcommand{\numMeanTestsAllProjects}{\num{39.2}\xspace}
\newcommand{\numMedianTestsAllProjects}{\num{3}\xspace}

% LOC per Project
\newcommand{\numMaxLocAllProjects}{\num{1.68} million\xspace}
\newcommand{\numMeanLocAllProjects}{\num{2791.9}\xspace}
\newcommand{\numMedianLocAllProjects}{\num{682}\xspace}
\newcommand{\numLocAllProjects}{\num{62} million\xspace}

% -- RQ 3

% a.)
\newcommand{\numRerunsOneNODTestNinetyFive}{\num{170}\xspace}
\newcommand{\numRerunsOneODTestNinetyFive}{\num{31}\xspace}

% b.)

\newcommand{\ratioNodFlakyTestsFoundAfterTenReruns}{\SI{33}{\percent}\xspace}
\newcommand{\ratioODFlakyTestsFoundAfterTenReruns}{\SI{54}{\percent}\xspace}

% c.)

\newcommand{\numRerunsEightyPercentNodFlakyTestsFivety}{\num{110}\xspace}
\newcommand{\numRerunsEightyPercentNodFlakyTestsNinetyFive}{\num{472}\xspace}

\newcommand{\numRerunsEightyPercentODFlakyTestsFivety}{\num{49}\xspace}
\newcommand{\numRerunsEightyPercentODFlakyTestsNinetyFive}{\num{209}\xspace}

% d.)

\newcommand{\numRerunsUnveilNODFlakyFailureNinetyFive}{\num{1}\xspace}
\newcommand{\numRerunsUnveilODFlakyFailureNinetyFive}{\num{3}\xspace}

% -- Meta-Data-Analysis

% Topic
\newcommand{\numAllProjectsWithTopic}{\num{7480}\xspace}
\newcommand{\numAllProjectsWithoutTopic}{\num{14872}\xspace}

\newcommand{\numAllTestsWithTopic}{\num{338716}\xspace}
\newcommand{\numAllTestsWithoutTopic}{\num{537470}\xspace}

\newcommand{\numFlakyProjectsWithTopic}{\num{339}\xspace}
\newcommand{\numFlakyProjectsWithoutTopic}{\num{667}\xspace}

\newcommand{\numNODFlakyProjectsWithTopic}{\num{92}\xspace}
\newcommand{\numNODFlakyProjectsWithoutTopic}{\num{187}\xspace}

\newcommand{\numODFlakyProjectsWithTopic}{\num{222}\xspace}

% Dev Status
\newcommand{\numAllProjectsWithDevStatus}{\num{9703}\xspace}
\newcommand{\numAllProjectsWithoutDevStatus}{\num{12649}\xspace}

\newcommand{\numAllTestsWithDevStatus}{\num{435600}\xspace}
\newcommand{\numAllTestsWithoutDevStatus}{\num{440586}\xspace}

\newcommand{\numFlakyProjectsWithDevStatus}{\num{429}\xspace}
\newcommand{\numFlakyProjectsWithoutDevStatus}{\num{577}\xspace}

% Github stars
\newcommand{\numMaxStarsAllProjects}{\num{149185}\xspace}
\newcommand{\numMeanStarsAllProjects}{\num{117.9}\xspace}
\newcommand{\numMedianStarsAllProjects}{\num{4}\xspace}

% ---

% iDFlakies Java
% LOC per Project
\newcommand{\numMinLocAllJavaProjects}{\num{638}\xspace}
\newcommand{\numMaxLocAllJavaProjects}{\num{1.65} million\xspace}
\newcommand{\numMeanLocAllJavaProjects}{\num{79640.20}\xspace}
\newcommand{\numMedianLocAllJavaProjects}{\num{19733.5}\xspace}
\newcommand{\numLocAllJavaProjects}{\num{6.4} million\xspace}

% ---

% NOD categories

\newcommand{\numNODNetwork}{\num{42}\xspace}
\newcommand{\numNODRandom}{\num{37}\xspace}
\newcommand{\numNODIO}{\num{7}\xspace}
\newcommand{\numNODTime}{\num{4}\xspace}
\newcommand{\numNODAsyncWait}{\num{3}\xspace}
\newcommand{\numNODConcurrency}{\num{3}\xspace}
\newcommand{\numNODResourceLeak}{\num{2}\xspace}
\newcommand{\numNODTestCaseTimeout}{\num{1}\xspace}
\newcommand{\numNODUnorderedCollection}{\num{1}\xspace}
\newcommand{\numNODTooRestrictiveRange}{\num{0}\xspace}
\newcommand{\numNODPlatformDependency}{\num{0}\xspace}
\newcommand{\numNODTestSuiteTimeout}{\num{0}\xspace}

\begin{abstract}
%
% Motivation
  Tests that cause spurious failures without any code changes, i.e.,
  \emph{flaky tests}, hamper regression testing, increase maintenance
  costs, may shadow real bugs, and decrease trust in tests.
%
% Previous studies -> need for additional data-set
  While the prevalence and importance of flakiness is well
  established, prior research focused on Java projects, thus raising
  the question of how the findings generalize.
  In order to provide a better understanding of the role of flakiness
  in software development beyond Java, we empirically study the
  prevalence, causes, and degree of flakiness within software written
  in Python, one of the currently most popular programming languages.
%
% Study Setup
  For this, we sampled \numAllProjects open source projects from the
  popular PyPI package index, and analyzed their \numAllTests test
  cases for flakiness.
  Our investigation suggests that flakiness is equally prevalent in
  Python as it is in Java. The reasons, however, are different: Order
  dependency is a much more dominant problem in Python, causing
  \OdFlakyTestsOverAllFlakyTests of the \numFlakyTests flaky tests in our dataset.
  Another \InfrastructureFlakyTestsOverAllFlakyTests were caused by test infrastructure
  problems, which represent a previously undocumented cause of
  flakiness.  The remaining \NodFlakyTestsOverAllFlakyTests can mostly be attributed
  to the use of network and randomness APIs by the projects, which is
  indicative of the type of software commonly written in Python.
  Our data also suggests that finding flaky tests requires more
  runs than are often done in the literature: A \SI{95}{\percent}
  confidence that a passing test case is not flaky on average would
  require \numRerunsOneNODTestNinetyFive reruns.

\end{abstract}


\input{sections/10_intro}
\input{sections/20_background}
\input{sections/30_approach}

\input{sections/40_evaluation}
\input{sections/50_related}
\input{sections/60_conclusions}

%\section*{Acknowledgements}\label{sec:acknowledgements}
%
%We thank Wenzl and Schnitzel for their constant inspiration and support :)
%Many animals were harmed during the writing of this paper,
%but they tasted soooooo well :P

\balance
\bibliographystyle{IEEEtran}
\bibliography{related.bib}

\end{document}

%% file: defs.tex
\usepackage[UKenglish]{babel}

\usepackage{graphicx}
\graphicspath{{./img/}}
\usepackage{placeins}
\usepackage{subcaption}
\usepackage{enumitem}
\usepackage{balance}

\usepackage[%
  group-minimum-digits=4,%
  list-final-separator={, and },%
  add-integer-zero=false,%
  free-standing-units,%
  binary-units,%
  detect-weight=true,%
  detect-inline-weight=math,%
]{siunitx}
\usepackage{microtype}

\usepackage[colorinlistoftodos,prependcaption,textsize=tiny]{todonotes}

\usepackage{xspace}

\usepackage[capitalise]{cleveref}

\crefformat{footnote}{#2\footnotemark[#1]#3}

\usepackage{tabularx}

\newcommand{\toolname}[1]{%
  \textsc{#1}\xspace%
}

\newcommand{\summary}[2]{
    \vspace{1em}
    \noindent
    \colorbox{gray!20}{%
        \parbox{.97\linewidth}{%
                \textbf{Summary \textit{(#1)}}
            #2
        }%
    }%
}%

\usepackage{csquotes}
\usepackage{csvsimple}
\usepackage{listings}
\usepackage{xcolor}

\definecolor{codegreen}{rgb}{0,0.6,0}
\definecolor{codegray}{rgb}{0.5,0.5,0.5}
\definecolor{codepurple}{rgb}{0.58,0,0.82}
\definecolor{backcolour}{rgb}{0.95,0.95,0.92}

\lstdefinestyle{mystyle}{
  backgroundcolor=\color{backcolour},
  commentstyle=\color{codegreen},
  keywordstyle=\color{magenta},
  numberstyle=\tiny\color{codegray},
  stringstyle=\color{codepurple},
  basicstyle=\ttfamily\footnotesize,
  breakatwhitespace=false,
  breaklines=true,
  breakindent=10pt,
  captionpos=b,
  keepspaces=true,
  numbers=left,
  numbersep=5pt,
  showspaces=false,
  showstringspaces=false,
  showtabs=false,
  tabsize=2
}

%% file: sections/10_intro.tex
\section{Introduction}\label{sec:intro}

Regression testing is a widely adopted practice in modern software
development. When new code gets checked in to the version control
system, an automated testing pipeline is triggered ensuring that the
most recent changes did not break existing functionality.  The basic
assumption behind regression testing is that the tests themselves
behave deterministically.  If a test does not behave
deterministically, but passes and fails when run multiple times
without any changes to the code, the test is regarded as \emph{flaky}.
Flaky tests confront developers with a dilemma: If they continue
taking all test failures seriously, they may waste precious time and
resources trying to find bugs that might not even be in the system
under test~(SUT), but rather in the test code or in the test
infrastructure.  On the other hand, if they disable tests that show
flaky behaviour, they may reduce the effectiveness of their test suite
and may miss bugs.

The most common strategy to reduce the impact of flaky test failures
is to rerun tests upon failure multiple times and accept a single
passed execution to make the test qualify as passed.  Many test
frameworks and continuous integration systems support marking tests as
flaky and running them up to a specified number of times upon failure
before reporting them as actually failed. This practice, however, has
multiple drawbacks:
First, it can only mitigate the problem by making flaky failures less
likely, as a flaky test could still fail all reruns. For large test
suites with high flaky failure rates, this approach may not be
effective at making a build pass.
Second, it might hide problems that should actually be fixed.  While
flakiness is often suspected to be rooted in the test code, it might
also be caused by the code under test itself.  Rerunning flaky tests
might therefore mask actual bugs.
Third, it wastes resources.
%While cloud testing systems have given us
%access to much greater amounts of computational power, their
%capabilities are still limited.
Google, for example, reportedly spends \SI{2}{\percent} to
\SI{16}{\percent} of their resources on re-running flaky
tests~\cite{Micco2017}.
Consequently, there is a need to study flakiness in depth in order to
understand its nature and to devise strategies to avoid it.

Previous research aiming to provide a deeper understanding of flaky
tests revolves around a limited set of Java projects. While Java is
widely used, other programming languages gained huge popularity over
the last decade, in particular Python: More than \num{200000} packages
are listed in the Python Package Index~(PyPI)\footnote{https://pypi.org/, accessed 2021--01--18.} at the time of this
writing, ranging from web frameworks such as Django to data analysis
tools such as NumPy or machine-learning libraries such as TensorFlow.
Despite its huge popularity, very little is known about flakiness in
Python, whether previous findings on Java also apply to Python, and
what further research is required in order to mitigate the problem of
flakiness in the Python world.

In this paper, we aim to fill this gap
%. We investigate the problem of
%flaky tests
by conducting a large empirical study on \numAllProjects
Python projects, consisting of \numAllTests test cases. Using a total
of \num{400} re-runs of these tests, we shed light on the questions of
(1) how prevalent flaky tests are in Python, (2) what the root causes
of flakiness are in Python, and (3)~just \emph{how} flaky these flaky
tests really are.
In detail, this paper  makes the following contributions:
\paragraph{Dataset} We derive a large dataset of \numAllProjects
Python projects with \numAllTests tests, of which \numFlakyTests tests
from \numFlakyProjects projects show flakiness. The resulting dataset,
which we share with the community~\cite{gruber_martin_2021_4450435},
consists of all artifacts as well
as the data produced by \num{400} test runs.
\paragraph{Study} We evaluate the test results with regard to the
extent, the cause, and the degree of flakiness we observed. For each
flaky test we provide a classification for the root cause of its
flakiness, and we further manually investigate 100 non-order-dependent
flaky tests to additionally provide a fine-grained classification into
13 established categories.
%; we offer possible explanations for the
%patterns we recognized, and compare our results to existing studies
%
\paragraph{Methodology} Using the extensive amount of data on flaky
tests in Python, we derive a novel, more stable approach to
estimate the number of reruns needed in order to expose possible
flakiness at a specific confidence level.

% -- Teaser findings -> how do they help
%
Our study shows that flakiness is an equally prevalent problem in
Python as it has been shown to be in Java. The reasons for flakiness,
however, differ: order-dependency between tests is a much more
dominating reason in the context of Python than it is for Java, and
non-order-dependent tests are predominantly caused by network and
randomness APIs, which are representative of the common application
areas of Python. We also identify infrastructure flakiness as a new
type of test flakiness, which may in particular affect researchers
conducting large experiments on flaky tests.
By providing statistical estimates of the required reruns to detect or
mitigate flakiness, and by releasing all data freely, we hope to
foster research on test flakiness in Python, and on automated
identification and classification techniques for flakiness.
%

%%% Local Variables:
%%% mode: latex
%%% TeX-master: "../main"
%%% End:

%% file: sections/20_background.tex
\section{Background}\label{sec:background}

%In order to address flakiness,
Several approaches have been proposed to automate the identification, classification, and elimination of flaky tests.

\subsection{Types of Flakiness}%
\label{sec:types_of_flakiness}

Luo et al.~\cite{LHE+14} introduced 10 categories of flakiness (\emph{Async Wait}, \emph{Concurrency}, \emph{Test Order Dependency}, \emph{Resource Leak}, \emph{Network}, \emph{Time}, \emph{IO}, \emph{Randomness}, \emph{Floating Point Operations}, and \emph{Unordered Collections}), which were later extended by Eck et al.~\cite{EPC+19} with the categories \emph{Too Restrictive Range}, \emph{Test Case Timeout}, \emph{Platform Dependency}, and \emph{Test Suite Timeout}.
Unlike most other causes, order dependencies can be properly identified automatically~\cite{ZJW+14,GBZ18}. Therefore, these categories are often grouped into \emph{order-dependent} and \emph{non-order-dependent} causes (the latter referring to all other 13 categories).

% Why are order-dependencies of concern? (question by reviewer)
One might argue that failures caused by order dependencies can be easily avoided by enforcing a particular test order.
However, dependencies between test cases can still cause failures as adding new tests or removing existing ones might break the test suite.
Therefore, developers should always be interested in avoiding test dependencies.

% -- Problems caused by order-dependency
% 1.) rely on implicit and undocumented knowledge -> better use fixtures
% 2.) performance (no more parallelization possible)
% 3.) adding new tests or deleting existing ones might break cause failures.

% OD classification: victims and brittles
We can further categorize order-dependent tests as follows~\cite{SLO+19}:
A test~$t$ can be order-dependent either because another test~$p$ running before $t$ disturbs its execution, or because another test~$s$ is not run before $t$, although $t$ requires $s$ to run before it.
%
% victim
In the first case, $t$ is called a \emph{victim} and $p$ is called a \emph{polluter}.
Test $p$ changes a shared state that $t$ tries to read from in a way that $t$ fails.
When run in isolation, a victim passes, as the state is not affected by the polluter.
%
% brittle
In the second case, $t$ is called a \emph{brittle} and $s$ is called a \emph{state-setter}.
Test $t$ needs $s$ to set up a shared state, e.g., a database connection, before it can run successfully.
When run in isolation, a brittle fails if the required state has not 
been set up.

% \enquote*{
%     A victim is an order-dependent test that consistently passes when run by itself in isolation from other tests.
%     The reason why a victim fails in a failing test order is that there is at least one test that runs before the victim, and these tests `pollute' the state (e.g., global variable, file system, network) on which the victim depends.
%     We call such state-polluting tests polluters.
%     [\ldots]
%     In contrast to a victim, an order-dependent test is a brittle if the test consistently fails when run by itself in isolation.
%     Intuitively, because a brittle fails in isolation and yet has a passing test order, then its passing test order must contain one or more tests that set up the state for the brittle to pass.
%     We refer to a test order that sets up the state for a brittle as a state-setter.
% }~\cite{SLO+19}

%

% Infrastructure flakiness
Besides the already introduced categories, there exists a 15th category not previously discussed: \emph{infrastructure flakiness}.
% Definition
It describes a test being flaky due to reasons outside the project's code but inside the test execution environment, for example failing installation of dependencies.
Infrastructure flakiness differs from other types of flakiness as it is not caused by the project itself, but by external components. % used by the developer.
It is therefore a form of transitively induced flakiness.

% Example
An example for infrastructure flakiness we experienced is the \texttt{pip} Python package management tool failing to install certain dependencies, resulting in \texttt{ModuleNotFoundError}s or \texttt{FileNotFoundError}s when executing the tests.
Despite the cause of flakiness being the network, this cannot be classified as network flakiness, as it is not the project which is trying to access the network.
% Review critic: more on infrastructure flakiness
Other causes involve permission errors and a lack of disk space.
%
% References
So far, infrastructure flakiness has not been formalized, although its effects have been previously observed, for example in a previous study which mentions that, out of 315 tests showing flakiness in continuous integration, only 44 cases were reproducible locally using 100 reruns~\cite{LGN+19}.

\subsection{Detecting Flaky Tests}%
\label{sec:detecting_flaky_tests}

To study flakiness, researchers have applied two strategies to build
datasets containing flaky tests:
(1)~Search for commits fixing flakiness or issues reporting flaky tests in version control systems or bug trackers;
(2)~Rerun tests up to a certain number of times and check whether their verdicts change between runs.

Luo et al.~\cite{LHE+14}, who conducted one of the first studies on flaky tests, used the first method by mining \num{201} commits that likely fix flaky tests in \num{51} open-source projects and manually verified this assumption.
In order to obtain a larger set of flaky tests in an automated fashion, later studies employed the second method.
They used various numbers of reruns (cf.~\cref{tab:reference_repetition}) without investigating this number in detail.
By using \num{400} reruns, which is more than most prior studies, we aim to derive a proper estimation on how many reruns should be conducted to build a representative dataset of flaky tests.

\begin{table}[tpb]
    \centering
    \caption{
        Previous studies and their identification strategies
    }
    \label{tab:reference_repetition}
    \begin{tabular}{rl@{}r}
        \toprule
        Source      & identify flakiness via                                & no.\ repetitions \\
        \midrule
        ~\cite{LHE+14,TSM18} & search for commits fixing flaky tests                 & - \\
        ~\cite{EPC+19} & search for already fixed flaky tests                  & - \\
        ~\cite{ZJW+14} & search in issue tracker                               & - \\
        ~\cite{BLH+18} & rerun, same order                                     & 5 \\
        ~\cite{LOS+19} & rerun, different order                                & 20-60* \\
        ~\cite{SLO+19} & rerun, same order                                     & 10 \\
        ~\cite{LGN+19} & rerun, same order                                     & 100 \\
        ~\cite{SBM19}  & rerun, different order chosen by PIT~\cite{Coles2016} & 17 \\
        ~\cite{GBZ18}  & data-flow analysis**                                  & - \\
        \midrule
        \multicolumn{3}{l}{\footnotesize * \toolname{iDFlakies} reruns failing test orders up to two times.}\\
        \multicolumn{3}{p{0.95\linewidth}}{\footnotesize ** data-flow analysis can only be used to detect order-dependent flakiness.}\\
        \bottomrule
    \end{tabular}
\end{table}

% \todo{add recently read papers: Intermittently Failing Tests in the embedded systems domain, a study on the lifecycle of flaky tests, vocabulary of flaky tests}

% Papers that directly mention conducting 10 reruns to be a good idea:
% ~\cite{SLO+19}
% ~\cite{SBM19}

% related work uses similar datasets -> not representative
%
Most previous work focused on a limited number of popular and large Java projects; several studies~\cite{SLO+19,LOS+19,BLH+18} use similar datasets, originating in the study by Luo et al.~\cite{LHE+14}.
%
%The reason for this sampling practice lies in the purpose the chosen projects were meant serve:
%Most previous datasets was for the proposed approaches and tools to be validated against,
%, or to derive insights about flaky tests by interviewing developers.
%
%With that in mind,
% to sample projects that are already known to contain flaky tests:
This sampling approach
is reasonable for measuring the performance of a tool aiming to detect or classify flakiness, as it makes the evaluation more stable and comparable to other studies.
However, this practice does not contribute to the overall understanding of flakiness, especially not outside the Java-world, where much less research is available.
%
% Hence, while these results demonstrate the effectiveness of their tools, it remains an open question how they generalize to the multitude of projects and programming languages existing.
%
%Our study is targeting this gap:
%We examined all projects in the Python project index for test flakiness in order to paint a complete picture of the extent of flakiness present in modern software development.
% Besides that, our dataset can also be used in order to validate future tools and frameworks, just as any previously proposed dataset.

\subsection{Mitigating Flaky Tests}%
\label{sec:mitigating_flaky_tests}

To mitigate flakiness, researchers have proposed several techniques that aim at detecting flaky tests in a resource efficient and automated way, classifying flaky tests in order to assist the debugging process, or automatically fixing flaky tests.

% -- Detection

\toolname{DeFlaker}~\cite{BLH+18} and \toolname{iDFlakies}~\cite{LOS+19} are both tools for automatically detecting flaky tests, offering performance advantages over repetitive reruns.
\toolname{DeFlaker} does so by analyzing coverage information and test verdicts from prior runs, therefore completely avoiding re-executions of any tests.
\toolname{iDFlakies} aims to detect flaky tests by applying a smart random-order rerun strategy, allowing it to also partially classify the root cause of the flakiness. However, it can only distinguish between order-dependent and non-order-dependent flaky tests.

% -- OD exclusive
Several other techniques focus exclusively on order-dependent test:
\toolname{PraDet}~\cite{GBZ18} tries to detect order-dependent flaky tests by using data-flow analysis to filter test orders containing potential order-dependencies, minimizing the number of test runs needed to expose order-dependent behavior.
\toolname{iFixFlakies}~\cite{SLO+19} aims at automatically fixing order-dependent flaky tests by suggesting patches extracted from other test code.

% -- Classification

% Describe that Rootfinder
\toolname{Rootfinder}~\cite{LGN+19} aims to derive a more fine-grained classification for non-order-dependent flaky tests:
Using a binary instrumentation framework it tries to collect distinctive information about a test execution, which it then compares between passing and failing runs in order to find patterns that predict test failures.
%
% Latest google stuff
Another classification tool proposed by Ziftci et al.~\cite{Ziftci2020} can also help to identify the root cause of flakiness by comparing the execution of passing and failing runs and pointing at the first line of code where the two executions diverge. Both approaches provide support, but cannot fully automate, finding the root causes of flakiness.

% Simulation

% For a grained classification of flaky tests, \toolname{Rootfinder}~\cite{LGN+19} can be applied:
% Using a binary instrumentation framework it collects logs about the test execution, from which it then tries to extract distinctive information in the form of so-called predicates.
% Predicates assert, for example, the return value of a function against a constant or dynamic value, the order in which methods are called, or the time that a method takes to complete.
% Comparing the predicates of passing and failing runs, the tool tries to find patterns that seem to predict a test failure.

\subsection{Types of Flakiness  in Python}%
\label{sec:possible_causes}

The categorization of flaky tests discussed in
Section~\ref{sec:types_of_flakiness} is not language specific, and
thus also applies to Python. There are, however, some language
specific peculiarities that have an impact on some types of flakiness.

% Floating point
\noindent\textbf{Floating point flakiness:}
Flakiness in Python cannot be caused by floating point
operations, as floating point arithmetic in Python is---despite
suffering from the well-known binary-decimal representation issues of
IEEE-754---always deterministic.

% Platform dependency
\noindent\textbf{Platform dependency:}
In many ways, Python hides the underlying system structure from the user
for example by automatically extending the size of an integer in case its
value reaches the word limit.
Nevertheless, it is possible for an execution to differ because of platform
dependencies, for example because the size of an object
(which is accessible via \texttt{sys.getsizeof})
differs between 32-bit and 64-bit systems.

% Unordered collection
\noindent\textbf{Unordered collections:}
The category \emph{Unordered collection} plays a special role in
Python: The internal ordering within a set depends on the
\texttt{\_\_hash\_\_} function.  This function used to be
deterministic up until Python~3.2 from where on it was randomly
seeded making it non-deterministic, a change made due to security
reasons. After all, this means that Python (in its default
configuration) only has unordered collections since Python~3.3.  It
is worth noting that the default behavior of collections is subject to frequent change in Python, as for example dictionaries are now
order-preserving since Python~3.8.

% Any category is possible, except for randomness
%Overall we were able to create a minimal Python example for all the 14 categories of flakiness, except for floating point operations.

%%% Local Variables:
%%% mode: latex
%%% TeX-master: "../main"
%%% End:

%% file: sections/30_approach.tex
\section{Study Setup}\label{sec:approach}

In this paper we aim to empirically answer the following research questions:
\begin{description}
  \item[RQ1:] How frequently does flakiness occur in Python?
% (possibly also cross-check against how many are labelled as such using the \texttt{flaky} library)
    %
  \item[RQ2:] What types of flakiness are prevalent in Python?
  \item[RQ3:] What is the degree of flakiness of flaky tests in Python?
\end{description}

\subsection{Dataset}%
\label{sec:dataset}

% -- origin and sampling of the projects
We scanned the entire Python Package Index~(PyPI) for suitable sample projects.
%
%PyPI
PyPI is the official third-party software repository for Python.
It features \num{284112} projects and \num{4787968} users (as of 2021--01--18), and is used as the default package source by many package managers including \texttt{pip}.

% Reviewer critic: Are all our projects also valid projects, not toy projects?
By using PyPI we hope to create a dataset, which is large and diverse enough to represent the language without including overly small toy projects.
Such toy projects will exist on GitHub, however, they are unlikely to be published to the community via PyPI.\@
While our dataset does contain small projects, the share of tests contributed by projects having less than 100 LOC is only \SI{3.5}{\percent}.

% Suitable Projects
We limited our exploration to projects whose source code is available on GitHub and whose tests can be executed using PyTest\footnote{\label{footnote:pytest}\url{https://pytest.org}, accessed 2021--01--18.}, the most commonly used test execution framework in Python.
%
% no. Projects
We ended up with \numAllProjects Python projects matching these criteria.
For each project, we consider its current state on 2020--08--16.

\begin{figure}[tpb]
\begin{subfigure}[b]{0.24\textwidth}
    \centering
    \includegraphics[width=\linewidth]{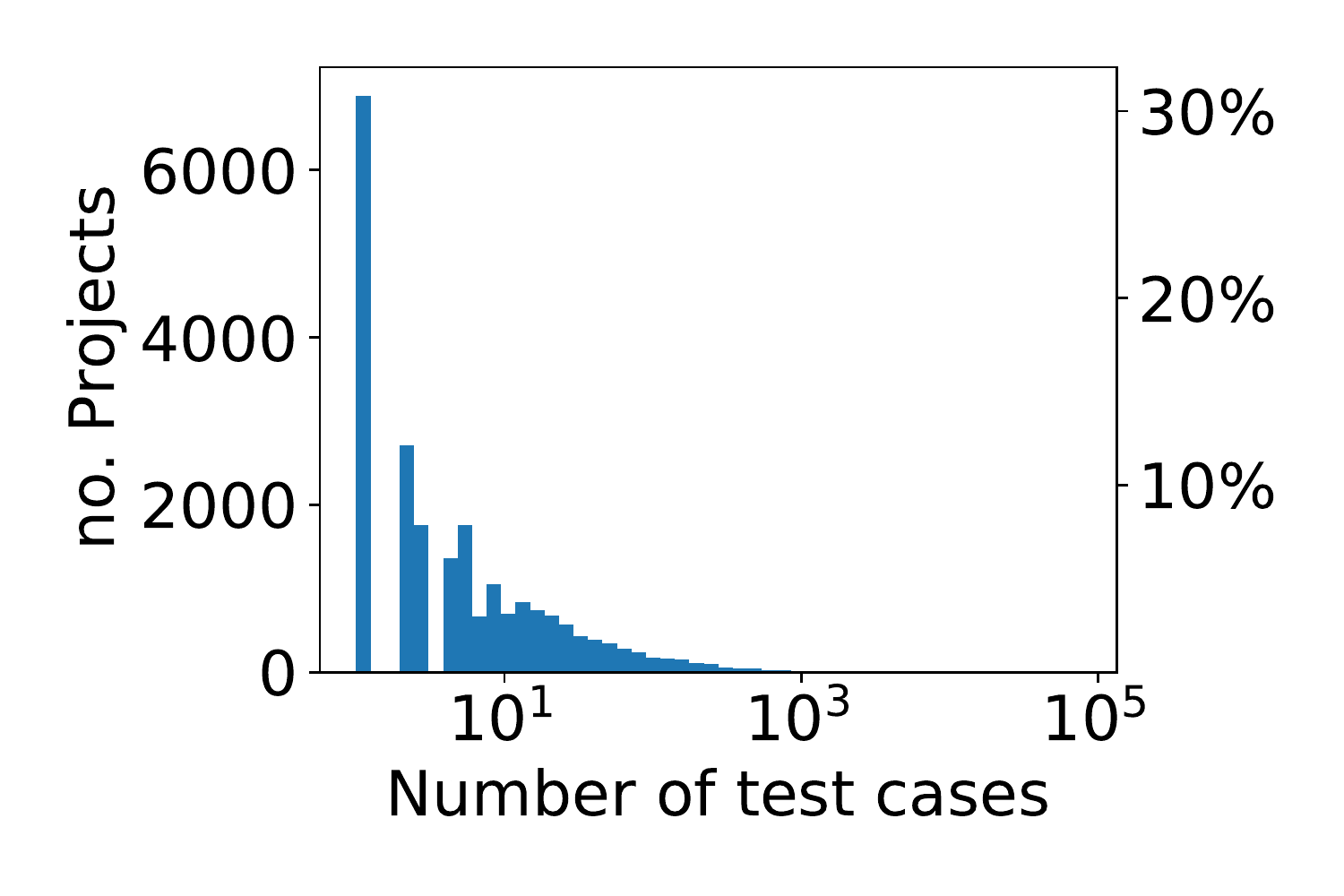}
    \caption{Number of tests}%
    \label{fig:no_tests_per_project}
\end{subfigure}
\hfill
\begin{subfigure}[b]{0.24\textwidth}
    \centering
    \includegraphics[width=\linewidth]{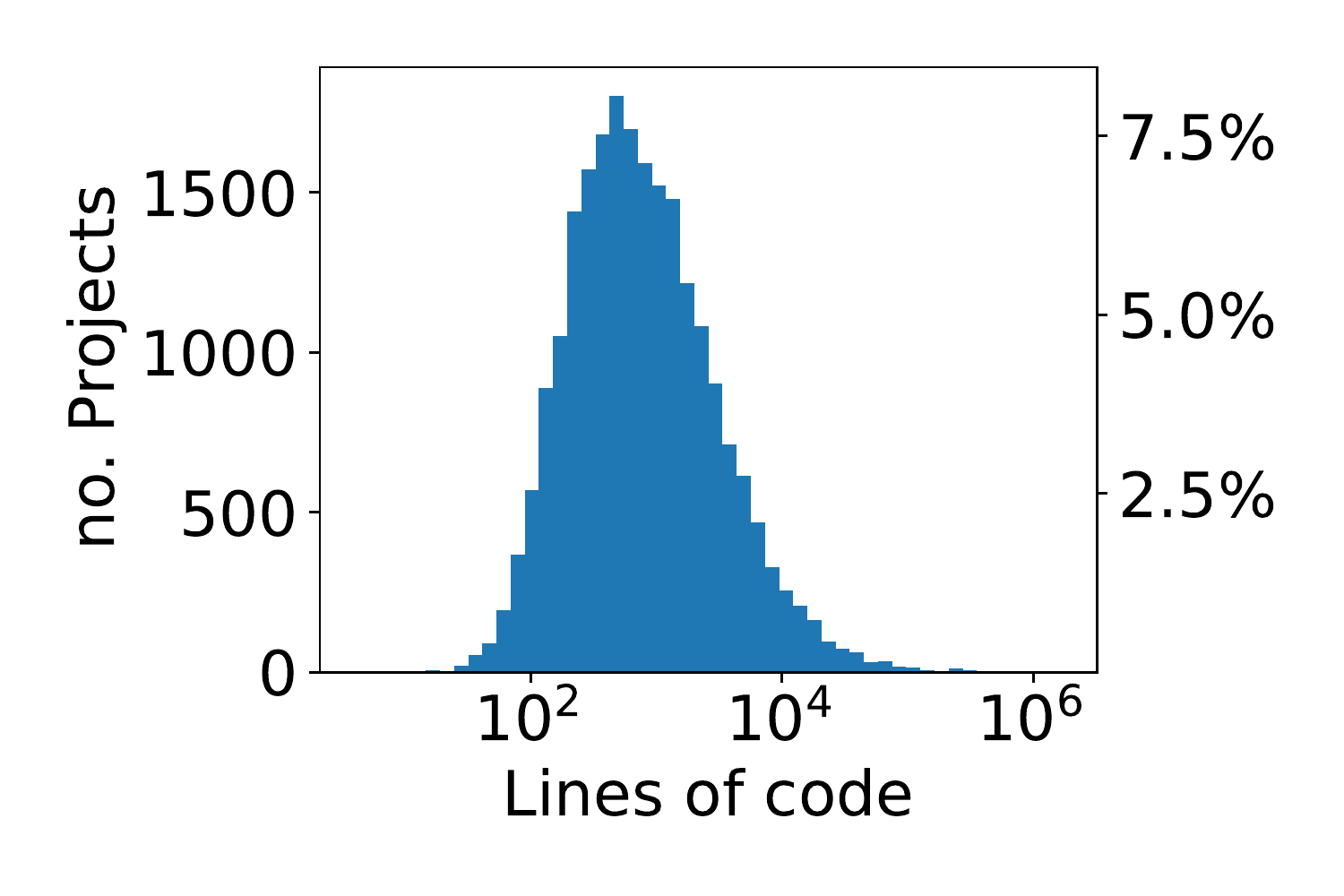}
    \caption{Lines of code}%
    \label{fig:loc_Python}
\end{subfigure}

\vskip\baselineskip

%-- Coverage
\begin{subfigure}[b]{0.24\textwidth}
    \centering
    \includegraphics[width=\linewidth]{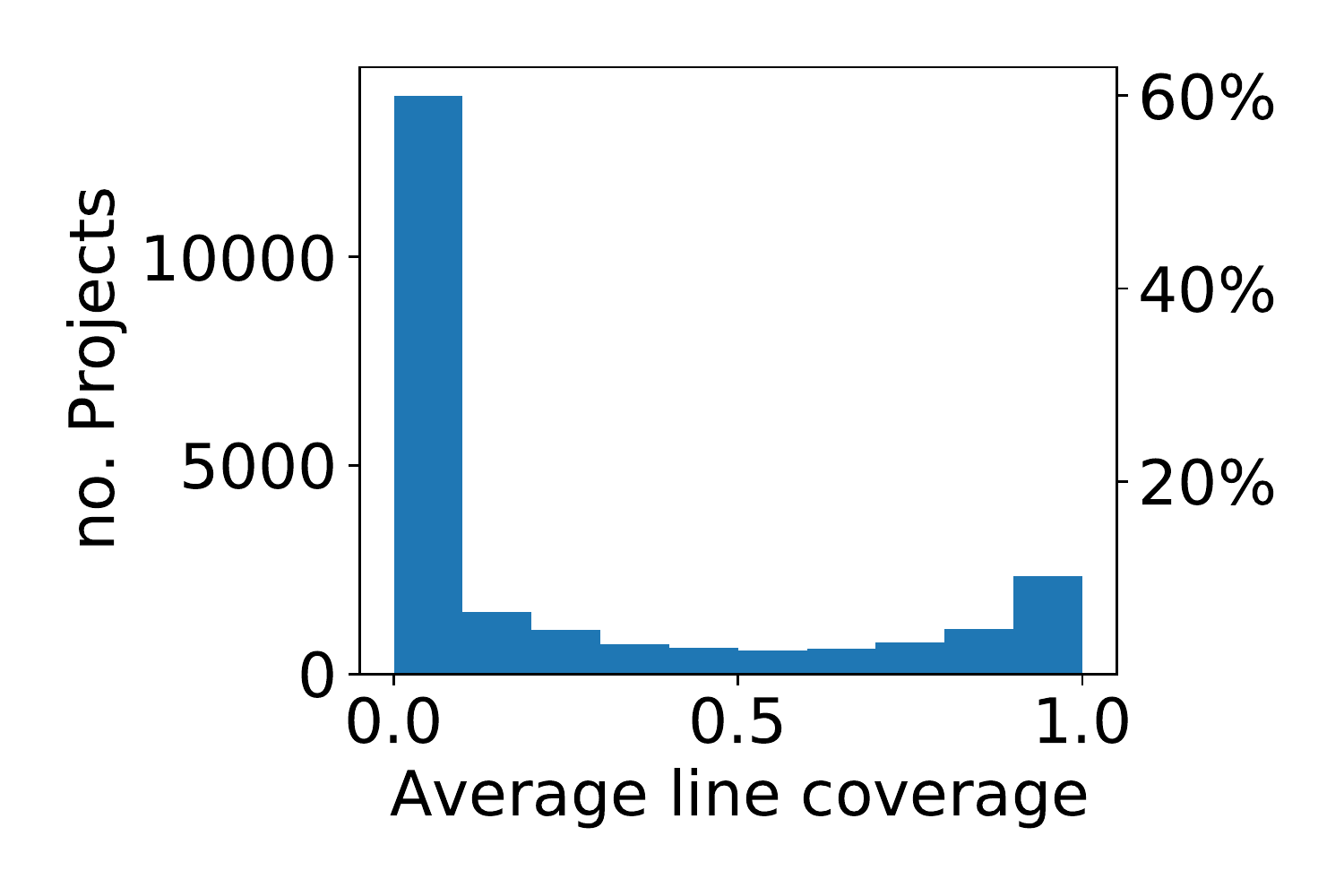}
    \caption{Average line coverage}%
    \label{fig:coverage}
\end{subfigure}
\begin{subfigure}[b]{0.24\textwidth}
        \centering
        \includegraphics[width=\linewidth]{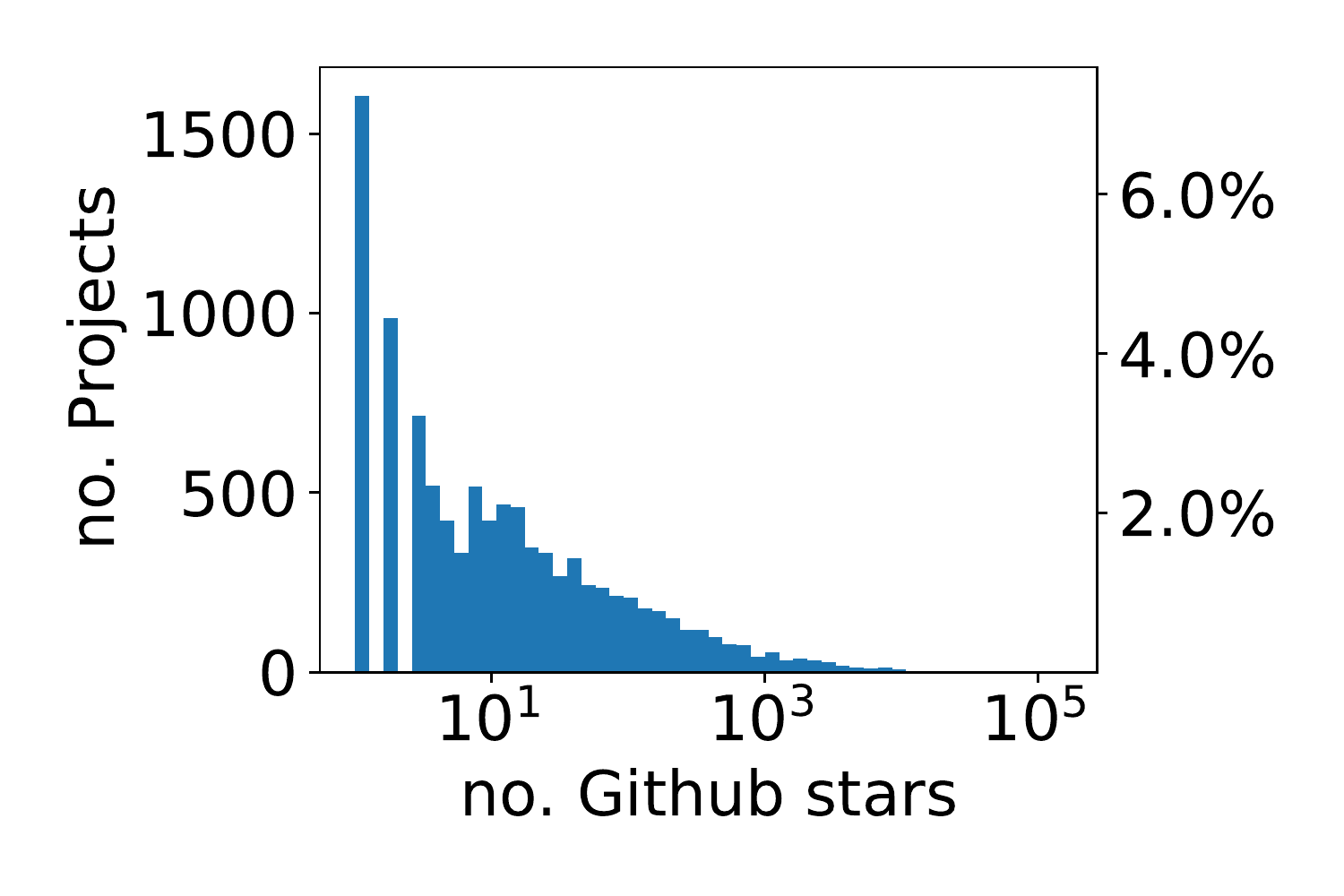}
        \caption{Number of Github stars}%
        \label{fig:stargazers_all}
\end{subfigure}
\caption{Statistics about the dataset of \numAllProjects Python projects}%
\end{figure}

% -- no. Tests
The projects contain between \numMinTestsAllProjects and \numMaxTestsAllProjects (project \enquote{capidup}) test cases.
The median number of test cases per project is \numMedianTestsAllProjects, the mean is at \numMeanTestsAllProjects.
In total the projects contain \numAllTests test cases.
\cref{fig:no_tests_per_project} shows a histogram of the distribution regarding the number of test cases per project.
To discover tests within a project, we rely on PyTest, which scans for files, classes, and methods, whose names contain the keyword \enquote{test}\footnote{\url{https://docs.pytest.org/en/stable/goodpractices.html#conventions-for-python-test-discovery}, accessed 2021--01--18.}.
%
% Parametrization
Python tests can be parametrized, meaning the same test is executed multiple times with different inputs.
PyTest reports each parametrization of a test as a separate test case.
This practice is reasonable as parametrized inputs are often of complex nature (e.g.\ files), covering different functionality within the code, and can therefore be considered separate tests.

% -- LOC
To estimate the sizes of our projects, we measured the non-comment source lines of code~(LOC) for each project using \toolname{cloc}~\cite{Danial2015}:
The smallest projects contain less than \num{10} LOC.\@
The largest project (\enquote{napalm-yang}) features \num{1.68} million lines of Python code.
The median number of LOC per project is \numMedianLocAllProjects, the mean is at \numMeanLocAllProjects.
The total number of LOC is slightly above 62 million.
\cref{fig:loc_Python} shows a histogram of the LOC-distribution.

As an indicator for the quality of the projects' test suites, we measured the average line coverage reported in our test runs, which is depicted in \cref{fig:coverage}.
Aside from a large number of low-coverage test runs (which can partly be attributed to error-ing tests), we also see about \SI{10}{\percent} of projects yielding a strong line coverage of above \SI{90}{\percent}.
The mean line coverage across all projects is \SI{24.6}{\percent}, the median is \SI{3.7}{\percent}.

% Popularity
To give an impression about the popularity of the investigated projects, \cref{fig:stargazers_all} depicts the number of Github stars per project.
The number of stars ranges from zero up to \numMaxStarsAllProjects (project \enquote{tensorflow}) with a median of \numMedianStarsAllProjects and a mean of \numMeanStarsAllProjects.

% Domain
We also wanted to know which domains of application the selected projects address, which we measured by looking at the topics developers assigned for their projects on PyPI.\@
PyPI provides a fixed set of 296 hierarchically organized topics, that developer can give their projects.\footnote{\url{https://pypi.org/classifiers/}, accessed 2021--01--18.}
%
% limitation: optional
This field, however, is optional, so not all projects have topics:
\numAllProjectsWithTopic of all investigated projects specified at least one topic.
%
% limitation: multiple + hierarchy not enforced
Furthermore, for some classifiers, multiple values of varying granularity can be specified without the hierarchy being enforced.
A single project can for example specify \enquote*{Topic :: Security :: Cryptography} without specifying \enquote*{Topic :: Security}.
While this is a technical possibility, it is common to specify all matching topics within the hierarchy to make the project more visible to search algorithms.
\cref{tab:all_topics} depicts the 25 most frequently assigned topics.
While the three top-most entries are very generic and offer little insights about the projects' domains, the following topics contain more information, showing that the selected projects cover a large variety of domains.
%, ranging from networking related topics over software development tools to system administration and text processing.

% Development status
At last, we looked at the maturity of the projects in the dataset, by examining their development status as specified on PyPI.
% Again, PyPI gives the developer the opportunity to specify this in form of the development status of a project, which can take one of seven values from planning to mature.
%
Of all investigated projects, \numAllProjectsWithDevStatus specified their development status.
We depict the number of projects per status in \cref{tab:all_dev_status}, showing a diverse picture with a tendency towards pre-production states.
%

% our dataset is representative
% Overall, these metrics show that our dataset constitutes a diverse set of Python projects, featuring projects that range from big to small, popular to less popular, heavily tested to sparsely tested, pre-alpha to production, coming from various domains.
% \gfdone{We cannot conclude whether it is representative. Either you say it's \emph{diverse}, or we simply drop this paragraph}

\begin{table}[]
        \caption{25 most used topics of the investigated projects}
        \label{tab:all_topics}
\resizebox{\columnwidth}{!}{%
\begin{tabular}{lr}
    \toprule
        Topic & no.\ Projects \\
    \midrule
         Software Development :: Libraries :: Python Modules & \num{2116} \\
                                                   Utilities & \num{1289} \\
                           Software Development :: Libraries &        910 \\
                                      Scientific/Engineering &        796 \\
                                        Software Development &        623 \\
                         Software Development :: Build Tools &        439 \\
                             Software Development :: Testing &        435 \\
                                        Internet :: WWW/HTTP &        420 \\
                     Internet :: WWW/HTTP :: Dynamic Content &        312 \\
                   Scientific/Engineering :: Bio-Informatics &        261 \\
           Scientific/Engineering :: Artificial Intelligence &        259 \\
                       Scientific/Engineering :: Mathematics &        238 \\
              Scientific/Engineering :: Information Analysis &        208 \\
                           Scientific/Engineering :: Physics &        205 \\
                                                    Internet &        171 \\
                                                    Database &        160 \\
                   Software Development :: Quality Assurance &        148 \\
                            System :: Systems Administration &        143 \\
 Software Development :: Libraries :: Application Frameworks &        133 \\
                     Scientific/Engineering :: Visualization &        123 \\
                                        System :: Networking &        121 \\
                               Text Processing :: Linguistic &        120 \\
                                                    Security &        114 \\
                                             Text Processing &        114 \\
                         Scientific/Engineering :: Astronomy &         99 \\
                                                UNSPECIFIED  &  \numAllProjectsWithoutTopic \\
    \bottomrule
\end{tabular}
}
\end{table}

\begin{table}[]
        \caption{development status of the investigated projects}%
        \label{tab:all_dev_status}
        \centering
\begin{tabular}{lr}
        \toprule
        Development status &  no.\ Projects \\
        \midrule
              4 - Beta & \num{3173} \\
             3 - Alpha & \num{2566} \\
 5 - Production/Stable & \num{2281} \\
         2 - Pre-Alpha & \num{1466} \\
          1 - Planning &        178 \\
            6 - Mature &         30 \\
          7 - Inactive &         12 \\
           UNSPECIFIED &   \numAllProjectsWithoutDevStatus  \\
        \bottomrule
\end{tabular}
\end{table}

\subsection{Detecting Flaky Tests}%

% Tool
To automate the detection of flaky tests, we built the \ourtool tool, which is  implemented in Python and is available under the GNU LGPL Open Source license.
We provide all source code of our analysis tool on GitHub\footnote{%
  \url{https://github.com/se2p/FlaPy}, accessed 2021--01--18
  }.

% --- Describe flaky-analysis tool
%
\ourtool takes as input the project folder of a Python project and executes the tests of the project either in a constant order between the test runs or in a random order.
The random-order execution allows choosing the level of granularity (i.e., class, module, package, or project level), at which the tests are shuffled.
Further, one can choose the number of repeated test runs that shall be executed.
At the core of our tool is PyTest\cref{footnote:pytest}, which collects all tests within the project directory and executes them.
%
% Python's heterogeneous dependency management landscape posed a special challenge in building the tool:
% As there is no universal standard for dependency management in Python, we build a heuristic that searches for requirements within the project's files and installs them to a virtual environment.
% Specifically, we search for files called ``requirements.txt'', with optionally ``dev'' or ``test'' prepended or postpended.
% By convention, each line within such files contains one dependency in a format that can be interpreted by \texttt{pip}.
% Furthermore, we look for a ``Pipfile'', which we parse using the equally named Python utility.
% We bundle all dependencies found and install them via \texttt{pip}.
% We do so for every test run, so each run is executed in a fresh virtual environment, which contains only the dependencies of the particular project.
%
We export the test results along with coverage information.
\ourtool uses \toolname{BenchExec}~\cite{BLW15} to separate the execution of the test cases from the parent process of our tool and to control the tests' access to external resources, such as network or hard disk.

\subsection{Analyzing Flaky Tests}%
\label{sec:analyzing_flaky_tests}

% \toolname{Rootfinder}~\cite{LGN+19} has originally been proposed to
% support the classification of flaky tests, but it is not usable for
% Python projects: The instrumentation framework it uses is not publicly
% available, and its requirements towards the instrumentation
% (especially in regard to timing) cannot be satisfied by an
% alternative, Python native, solution.
%
In order to support the classification of flaky tests, we developed an approach which searches the execution-traces of a test for a set of keywords, that are indicative for different categories of flakiness.
To determine relevant keywords, we created a
minimal representative Python test for each known category of flakiness and
traced multiple executions of these while collecting all called
functions using Python's tracing API.
%
% In order to collect the traces, we utilize Python's builtin \cmdtool{sys.setprofile} method, a functionality commonly used by debuggers.
% As for tests whose root cause of flakiness is known, we use the minimal examples introduced in Section~\ref{sec:possible_causes}.
% %
% Figure~\ref{fig:lst:trace} shows an excerpt from the trace of the test case \texttt{test\_network\_remote\_connection\_failure}.
%
% \begin{figure}[tpb]
%     \centering
%     \begin{lstlisting}[]
% --> ('test_flaky', '', 'test_network_remote_connection_failure')
% ----> ('requests.api', '', 'get')
% ------> ('builtins', 'dict', 'setdefault') #builtin
% <------ ('builtins', 'dict', 'setdefault') #builtin
% ------> ('requests.api', '', 'request')
% --------> ('requests.sessions', 'Session', '__init__')
% ----------> ('requests.utils', '', 'default_headers')
% ------------> ('requests.utils', '', 'default_user_agent')
% <------------ ('requests.utils', '', 'default_user_agent')
% ------------> ('builtins', 'str', 'join') #builtin
% <------------ ('builtins', 'str', 'join') #builtin
%               ...
%     \end{lstlisting}
%     \caption{Example of a trace for the test case in Fig.~\ref{fig:lst:minimal_flaky_test}.}%
%     \label{fig:lst:trace}
% \end{figure}
By looking at differences between these traces, we identified the
function calls that are most characteristic for the specific type of
flakiness.
\cref{tab:keywords_categories} shows the keywords we extracted from the traces of our minimal examples respective to each category for which we were able to find distinctive keywords.
% Table~\ref{tab:keywords_categories} therefore also contains non-deterministic APIs within Python together with their respective category of flakiness.

\begin{table}[tpb]
    \caption{Python APIs and keywords indicating flakiness}%
    \label{tab:keywords_categories}
    \centering
\begin{tabularx}{\columnwidth}{l>{\hangindent=2em}X}
    \toprule
        Category             & Keyword \\
    \midrule
        Async wait           & \texttt{sleep} \\
        Concurrency          & \texttt{thread} \\
                             & \texttt{threading} \\
        % Resource Leak        & \\
        IO                   & \texttt{builtins.stat} \\
                             & \texttt{pathlib.Path.is\_dir} \\
        Network              & \texttt{requests} \\
        Time                 & \texttt{time} \\
        Random               & \texttt{random} \\
        % Floating Point       & \\
        Unordered Collection & \texttt{\_\_hash\_\_} \\
                             & \texttt{builtins.set.\_\_contains\_\_} \\
    \bottomrule
\end{tabularx}
\end{table}

% \begin{table*}[htpb]
%     \centering
%     \caption{Keyword-trace-matrix \mgtodo{fit to width, maybe leave out empty columns + only mention Test funcname}}%
%     \label{tab:keywords_trace_matrix}
%     \csvautotabular[respect all,separator=semicolon]{tables/keyword_traces__chunk01.csv}
% \end{table*}

When aiming to classify a given flaky test, we can now trace the
execution of the flaky test, and search for appearances of the
representative flaky APIs (or keywords) within these traces. If one of
the keywords matches, this suggests that the test might be flaky due
to the corresponding category. If a keyword does not appear, that
category is unlikely.

This techniques suffers from two obvious limitations:
(1)~There are multiple ways to trigger the same type of flakiness.
We mitigate this issue by considering execution traces as well as the tests' code, enabling us to find indirect usages of our minimal flaky APIs.
(2)~In case multiple flaky APIs are used, we have no way of telling which caused the test to flake.
This, however, is a general limitation of a text-based search, in contrast to a semantic program analysis.
In consequence, we use this technique only to support a manual classification.

\subsection{Methodology}%

\subsubsection{RQ1: Prevalence of Flakiness in Python}%
\label{sec:study_setup_rq1_prevalence_of_flakiness_in_python}

% Methodology
In order to identify flaky tests in Python projects, we execute the
tests of a project in an isolated manner using
\ourtool.  For each project, we executed its tests
\num{200} times in the same order and \num{200} times in a random
order, shuffled on project level to expose as many order dependencies as possible.
% Execution setup
We performed the test executions on a \toolname{SLURM}-managed cluster~\cite{yoo2003slurm} consisting of 91 nodes giving each test job \SI{16}{\giga\byte} of RAM as well as a \SI{24}{\hour} timeout.
In total, this took \SI{484}{\hour}.

% Runtime `find 20200814_rerun -iname "results.tar.xz" -printf'%T+%p\n' | sort | less`
%
% 20200814_rerun: 2020-08-25+16:35 -- 2020-08-26+17:43 => ~25h
% 20200814_orig:  2020-07-27+23:24 -- 2020-08-08+21:57 => ~12d-1h = 287h
% 20200825_orig:  2020-09-23+14:02 -- 2020-09-26+10:19 => ~68h
% 20201006:       2020-10-01+21:28 -- 2020-10-06+05:49 => ~4d+8h = 96h+8h = 104h
%                                                         ------
%                                                         484h

% Chimaira01
% - Intel Xeon E5-2690 3Ghz
% - 20 cores, 64GB RAM

% Definition of flakiness
We consider a test to be flaky iff it passed at least once and failed
at least once.
%
% Topics
Inspecting the topics of the projects that contain flaky tests, we hope to
find an indication which domains are prone towards flakiness.
%
% Maturity
We also look at the maturity of projects containing flakiness,
expecting to find more flakiness in alpha- and beta-phase projects,
rather than in mature projects.

% frequency metric
% To measure the influence of a project's domain and maturity on flakiness,
% we calculate for each topic (respectively development status) the
% ratio between its frequency amongst projects containing at least one
% flaky test, and its frequency amongst all projects.
% %
% If this relative frequency is greater than $1$, the
% topic / development status is more common in projects
% containing flakiness than it is on average.
% If the relative frequency is smaller than $1$, it is rarer amongst projects
% containing flakiness.

% To measure the influence of a project's domain and maturity on flakiness, we compare the rate at which flakiness occurs amongst the projects of a certain topic (respectively development status) to the average rate at which a project contains flakiness.

\subsubsection{RQ2: Types of Flakiness in Python}%
\label{sec:study_setup_rq2_type_of_flakiness_in_python}

% --- Motivation
Besides investigating the mere extent of flakiness, we also want to provide explanations for the observed non-deterministic behavior, and therefore classify the flaky tests identified as part of RQ1.

% -- Infrastructure flakiness

% Infrastructure flakiness correlates with failure bulks
We noticed that infrastructure flakiness often appears in failure
bulks, where all runs executed on the same machine within a
short period of time fail.
%
% -> slice into iterations
In order to distinguish infrastructure flakiness from flakiness within the project, we therefore sliced the \num{200} runs into \num{10} iterations of \num{20} runs each.
All \num{20} runs within an iteration were executed on the same machine in an uninterrupted sequence.
The \num{10} iterations were distributed across different machines and always had several hours of temporal distance.
%
% Classify as infrastructure flaky
We consider a test to be flaky due to non-determinism in the test infrastructure, if it exhibits flaky behavior only between iterations but not within an iteration (e.g., the test passed for all runs within an iteration and failed for all runs within another).
In contrast to that, we consider a test to be flaky due to reasons lying within the project's code, iff it exhibits at least one passing and at least one failing run within the same iteration for at least one iteration.

% --- Categorization OD / NOD
For all non-infrastructure related flaky tests, we distinguish between order-dependent flakiness~(OD) and non-order-dependent flakiness~(NOD) as follows:
If a test shows flaky behavior in test runs featuring the same test order, it is being categorized as NOD, regardless of its behavior when executed in random order.
If a test shows no flaky behavior when run in the same test order but does show flaky behavior when executed in random order, it is categorized as OD.

% --- Categorization OD: victims / brittles
Following a previous approach~\cite{SLO+19}, we further categorize order-dependent tests by running them in isolation:
If an OD test always passes when run in isolation, it is a victim, if it constantly fails, it is a brittle.
%
% --- Categorizing NOD: manual classification + tracing
For NOD flakiness we furthermore distinguish the remaining 13 categories.
Unlike the classification of OD tests into victims and brittles, the classification of NOD tests cannot be easily automated, which is why we classify the NOD flaky tests manually.

% Sampling
As we found more NOD flaky tests than we could classify in a reasonable amount of time, we selected \num{100} of them via random-stratified sampling:
We randomly chose \num{100} projects out of the \numNODFlakyProjects projects that contained at least one NOD flaky test.
For each project we then randomly selected one of its NOD flaky test cases.
In doing so, we hope to retrieve an unbiased sample.
We specifically avoided picking multiple flaky tests from the same project, as they are likely to be flaky due to the same root cause.

% Manual classification
Each test case was then manually classified independently by two authors using
(1) the project's code,
(2) the test-execution reports from all \num{200} runs therefore including at least one failure-trace, and
(3) the category-distinctive keywords found by the keyword-trace-search (\cref{sec:analyzing_flaky_tests}).
%
%The keywords are supposed to give hints about which type of flakiness the author might be looking at and be especially useful in cases where the root cause of the flakiness is hidden within a third party library.
%
%Each author rated the confidence in his classification from 1~(sure) to 4~(unsure).
In the next stage, we resolved all cases in which the two authors came to a different conclusion or were both unsure regarding the category of flakiness, via an in-depth discussion.
%
% Aside from the category of flakiness, we also extracted the project domain and whether the flakiness as caused by using fixtures.

\subsubsection{RQ3: Degree of Flakiness}%
\label{sec:study_setup_rq3_degree_of_flakiness}

Besides root causes of flakiness, we also measure its degree. This is
mainly of interest to researchers aiming to derive representative
datasets on flakiness, who might be asking questions such as:
\begin{enumerate}[label=(\alph*)]
    \item \enquote*{How many times do I have to rerun a test to be \SI{95}{\percent} sure that the test is not flaky?}
    \item \enquote*{If I rerun my tests ten times, which portion of the existing flakiness can I expect to find?}
    \item \enquote*{How many reruns do I need in order to find
        % ($p$ percent of) all existing flakiness?
        \SI{80}{\percent} of all existing flakiness?}
%    \item \enquote*{In case a test execution fails, how many reruns should I conduct in order to be sure the failure indicates a bug and not just a flaky test?}
\end{enumerate}
From a practical point of view it is mainly relevant to understand how
often to rerun \emph{failing} tests to identify flakiness. Thus, for
practitioners a more relevant question might be:
\begin{enumerate}[label=(\alph*)]
    \item[(d)] \enquote*{In case a test execution fails, how many reruns should I conduct in order to be sure the failure indicates a bug and not just a flaky test?}
\end{enumerate}

% Besides conducting more reruns on a larger amount of projects than previous studies, we also
We propose an alternative metric for calculating the recommended number of reruns:
In order to estimate the number of reruns required to find flakiness, previous studies~\cite{BLH+18, SBM19} based their decision on the number of reruns needed to unveil flakiness \emph{once}.
For a test $t$ we call this number $n_{t,\mathrm{once}}$, which is defined in the following way:
Assume we run our tests $n$ times, resulting in a list of runs $\langle r_{1},r_{2},\ldots,r_{n}\rangle$ with the function $\verdict(t, r_i)$ defining the outcome of test $t$ in run $r_i$.
Following PyTest's JUnit-XML plugin\footnote{\url{https://docs.pytest.org/en/stable/usage.html#creating-junitxml-format-files}, accessed 2021--01--18.}, a test's verdict can take the values $\mathit{PASS}$, $\mathit{FAIL}$, $\mathit{ERROR}$, or $\mathit{SKIP}$.
$n_{t,\mathrm{once}}$ is the first index, for which the test $t$ showed both a passing and a non-passing ($\mathit{FAIL}$ or $\mathit{ERROR}$) execution:
\begin{align*}
    n_{t,\mathrm{once}} = & \max ( \min(\{ i\ |\ \verdict(t,r_i) = \mathit{PASS} \}), \\
                          & \min(\{ i\ |\ \verdict(t,r_i) \in \{ \mathit{FAIL}, \mathit{ERROR} \} \}))
  \end{align*}
Exposing all flaky tests within a test suite $T$ therefore requires
$max( \{ n_{t,\mathrm{once}}\ |\ t \in T \} )$ reruns.

There are two issues with this approach:
(1) it is hard to reproduce, as it is based on single-time events, which might have been an ``(un)lucky punch'';
(2) it is unstable, as by design it utilizes only a limited amount of data which can hardly be extended by conducting more reruns---all verdicts seen after the flakiness was exposed once are ignored.

We therefore propose a new method for estimating the required number of reruns to expose flakiness, which is based on all verdicts the tests exhibited, and is able to provide a confidence level:
For every test~$t$ we calculate its passing rate~$P_t(\mathit{PASS})$ as well as its non-passing rate~$P_t(\mathit{FAIL/ERROR})$ as the ratio between the number of passed (respectively failed or errored) executions, and the number of all executions.
%
% \begin{align*}
%     P_t(\mathrm{pass}) = \frac{|\{ t\ |\ \verdict(t,r_i)=\mathit{PASSED} \land 0<i<n\}|}{n}\\\\
%     P_t(\mathrm{fail}) = \frac{|\{ t\ |\ \verdict(t,r_i)=\mathit{FAILED} \land 0<i<n\}|}{n}
% \end{align*}
%
Note that these may not add up to \num{1.0}, as a test can also result in the verdict $\mathit{SKIP}$.\@
Assuming the executions of the same test in different runs are independent from each other, these rates are also the probabilistic chances for the test to pass/not pass its next execution.
The independence assumption is strengthened by the fact that we already filtered out bulked failures that occurred due to infrastructure flakiness.

We define the probability~$U_t(n)$ for unveiling the flakiness of test $t$ after $n$ reruns as one minus the probability of not seeing any flakiness, meaning the test never fails, never passes, or is always skipped (meaning it neither passes nor fails):
% \begin{align}
%       & P_t(\textrm{unveil flakiness after $n$ reruns}) \\
%     = & P_t(\textrm{at least 1x passed and 1x failed}) \\
%     = & 1 - P_t(\textrm{never failed = only passed/error/skip}) \\
%       &   - P_t(\textrm{never passed = only failed/error/skip}) \\
%       &   + P_t(\textrm{never passed or failed = only error/skip}) \\
%     = & 1 - {(1-P_t(\textrm{fail}))}^n \\
%       &   - {(1-P_t(\textrm{pass}))}^n \\
%       &   + {(1-P_t(\textrm{pass})-P_t(\textrm{fail}))}^n
% \end{align}
\begin{align*}
    U_t(n) = 1 & - (1 - P_t(\mathit{FAIL/ERROR}))^n \\
               & - (1 - P_t(\mathit{PASS}))^n \\
               & + (1 - P_t(\mathit{PASS}) - P_t(\mathit{FAIL/ERROR}))^n
\end{align*}

We then calculate $n_{t,p}$ as the minimum number of reruns needed to unveil the flakiness of test $t$ with a probability $p$, calling it the \emph{statistical number of reruns} with confidence $p$:
\begin{align*}
    n_{t,p} = \min(\{n\ |\ U_t(n) > p \})
\end{align*}

This metric addresses both issues of $n_{t,\mathrm{once}}$, as it is not based on single-time events and utilizes data from all reruns.
Question (a) can now be answered by calculating the median value of $n_{t,0.95}$ (\SI{95}{\percent} confidence) for all tests $t$ within a given dataset.
We visualize our results and answer questions (b) and (c) by looking at the sum of flaky tests found after $n$ reruns, suggested by metric $x$ for a set of tests $T$:
\begin{align*}
    S(n,x,T) = \vert \{ t\ |\ n_{t,x} \leq n\ \land\ t \in T \} \vert
\end{align*}
%
% We calculate $S$ for metric
% (1) $x = \mathrm{once}$, which is the number of flaky tests we found in our concrete experiment,
% (2) $x = 0.5$, which is the average number of flaky tests one can expect to find, as well as
% (3) $x = 0.95$, which is a more pessimistic estimation.
% %
% For the set of considered tests $T$, we differentiate between order-dependent flaky tests $T_{\mathrm{OD}}$ and non-order-dependent tests $T_{\mathrm{NOD}}$.
%
% Practitioner's question
To answer question (d), we calculate the chance of a failing flaky test to pass at least once within the next $n$ iterations, which is $1 - (1 - P_t(\mathit{PASS}))^n$, and derive $n$ such that this probability exceeds \SI{95}{\percent}.
We answer RQ3 separately for OD and NOD flaky tests, as for OD flaky tests repeated execution is only one way to expose their flakiness~\cite{GBZ18}, while for NOD flaky tests, there is currently no alternative to repetitive reruns. % are currently the only way to unveil their flakiness.

\subsection{Threats to Validity}\label{sec:threats_to_validity}

%Although we carefully conducted this research, several threats to validity of this study exist.

\textbf{Internal Validity.\quad} Python execution tracing has limitations; for example, some builtin functions cannot be traced, hence we potentially miss these calls in our traces during the keyword-trace-search (\cref{sec:analyzing_flaky_tests}).
However, two authors manually classified all cases independently, and the traces were only one source of input.
%and held extensive discussions in cases where the classifications of the authors did not match or the confidence of the classification was weak.
% Platform dependency
While we did execute the tests on different machines, these were still very similar both in their hardware- and their software-configuration,
% \todo{MG: platform-dependency might be misclassified as infrastructure flakiness}
which might have had a negative effect on the number of platform-dependency related flakiness we were able to expose.

\textbf{External Validity.\quad} We conducted our analysis on a large sample of projects from PyPI.\@
However, our sampling procedure resulted in \numAllProjects projects, which is only about \SI{10}{\percent} of all available projects on PyPI.
Therefore, our conclusions might not generalize to other Python projects, and they also might not generalize to other projects in other programming languages.
%, although our results on the amount of flakiness partially overlap with those from previous research on Java projects.

\textbf{Construct Validity.\quad} The number of iterations necessary to expose all cases of flakiness is unknown.
By using \num{400} reruns, we achieve a fairly high confidence that we exposed a large percentage of flaky tests.
Furthermore, by introducing a statistical metric we are able to report a necessary number of runs to expose flakiness with a given confidence level.
However, it is still possible that we misclassified non-order-dependent flaky tests as order-dependent, if they showed flakiness only in random order executions.

%% file: sections/40_evaluation.tex
\section{Results}\label{sec:evaluation}

\subsection{RQ1: Prevalence of Flakiness in Python}%
\label{sec:evaluation_rq1_prevalence_of_flakiness_in_python}

% How much flakiness did we find?
In total, we found \numFlakyTests tests exhibiting non-deterministic behavior in \numFlakyProjects projects.
This gives us a ratio of \FlakyTestsOverAllTests of all investigated tests being flaky and a ratio of \FlakyProjectsOverAllProjects of all investigated projects containing at least one flaky test.
% These numbers are also mentioned in Table~\ref{tab:no_flaky_projects_and_tests}.\todo{Table lacks a headline for at least the second and third column}

% Maturity
\cref{tab:flaky_dev_status} shows projects containing flakiness grouped by their maturity.
Against our expectation, flakiness is more common in projects with higher levels of maturity, than those in earlier development phases:
\SI{5.3}{\percent} of all stable projects contain flakiness whereas only
\SI{3.4}{\percent} of all pre-alpha projects do so.
This observation can be explained by the fact that projects in later phases in general do more testing (which can be seen in the last column of \cref{tab:flaky_dev_status}) and are therefore more likely to also contain at least one flaky test.

% Domain
\cref{tab:flaky_topic} depicts the flaky projects by topic.
We can observe a tendency towards the science and engineering domain, in particular towards artificial intelligence.
In contrast to that, build tools, libraries, and internet-related projects seem to contain less flakiness than average projects.
Unlike for the development status, however, these differences can not all be attributed to certain topics conducting more tests:
Projects specifying the topic with the highest flakiness rate have a
comparable amount of tests to projects which do not specify a topic,
suggesting that the topic does have direct influence on the prevalence
of flakiness within a project.
%
% The number of unspecified projects serves as a control variable:
% It shows that projects that do not specify a topic (or development status) do not contain flakiness more or less often than the average project, as their flakiness rate is very close to \SI{4.4}{\percent}, which is also the flakiness rate of all investigated projects.
% \todo{MG: maybe leave out paragraph about unspecified?}
%
% Removed outliers
Note that for \cref{tab:flaky_topic} we show only topics which occur at least 15 times in all flaky projects.
%

% \showthe\textwidth %> 506.295pt
% \showthe\columnwidth %> 241.14749pt

\summary{RQ1: How frequently does flakiness occur in Python?}{We found \numFlakyTests tests that exhibit flaky behavior, making up a portion of \FlakyTestsOverAllTests of the tests we examined, with flakiness being more common in more mature projects as well as projects from the scientific and engineering domain.}

% TODO maybe mention project with
% - the most flaky tests
% - the most flaky tests compared to the total number of tests
% - the most flaky tests per LOC

% \begin{table*}
%     \centering
%     \caption{Development Status of flaky projects}%
%     \label{tab:flaky_dev_status}
%     \begin{tabular}{lrlrll|r}
%         \toprule
%             Development Status &
%             \begin{tabular}{@{}r@{}}no.\ flaky\\projects\end{tabular} &
%             \begin{tabular}{@{}l@{}}frequency in\\flaky projects\end{tabular} &
%             \begin{tabular}{@{}r@{}}no.\\projects\end{tabular} &
%             \begin{tabular}{@{}l@{}}frequency in\\all projects\end{tabular} &
%             \begin{tabular}{@{}l@{}}rel.\ frequency\\flaky / all\end{tabular} &
%             \begin{tabular}{@{}r@{}}average no.\ tests\\per projects\end{tabular} \\
%         \midrule
%  5 - Production/Stable &    121 &  0.288783 &       2281 &   0.235082 &        1.228435 &           68.536168 \\
%              3 - Alpha &    113 &  0.269690 &       2566 &   0.264454 &        1.019797 &           18.260327 \\
%               4 - Beta &    131 &  0.312649 &       3173 &   0.327012 &        0.956078 &           35.148440 \\
%          2 - Pre-Alpha &     49 &  0.116945 &       1466 &   0.151087 &        0.774023 &           12.972715 \\
%         UNSPECIFIED           & \numFlakyProjectsWithoutDevStatus & \freqFlakyProjectsUnspecifiedDevStatus & \numAllProjectsWithoutDevStatus & \freqAllProjectsUnspecifiedDevStatus & \relFreqFlakyAllUnspecifiedDevStatus & \\
%         \bottomrule
%     \end{tabular}
% \end{table*}

\begin{table}
    \centering
    \caption{Development status of flaky projects}%
    \label{tab:flaky_dev_status}
    \begin{tabular}{lrrrr}
        \toprule
            Development status &
            \begin{tabular}{@{}c@{}}no.\ flaky\\projects\end{tabular} &
            \begin{tabular}{@{}c@{}}no.\\projects\end{tabular} &
            \begin{tabular}{@{}c@{}}flakiness\\rate\end{tabular} &
    \begin{tabular}{@{}r@{}}average\\no.\ tests\\/ project\end{tabular} \\
        \midrule
5 - Production/Stable  & 124                               & \num{2281}                      & \SI{5.3}{\percent} & 79.5 \\
        UNSPECIFIED    & \numFlakyProjectsWithoutDevStatus & \numAllProjectsWithoutDevStatus & \SI{4.5}{\percent} & 34.8 \\
             3 - Alpha & 116                               & \num{2566}                      & \SI{4.4}{\percent} & 22.5 \\
              4 - Beta & 133                               & \num{3173}                      & \SI{4.1}{\percent} & 53.9 \\
         2 - Pre-Alpha & 50                                & \num{1466}                      & \SI{3.4}{\percent} & 14.5 \\
          1 - Planning & 6                                 & \num{178}                       & \SI{3.4}{\percent} & 18.8 \\
        \bottomrule
    \end{tabular}
\end{table}

\begin{table}
    \centering
    \caption{Topics of flaky projects}%
    \label{tab:flaky_topic}

\resizebox{\columnwidth}{!}{

    % \begin{tabularx}{\columnwidth}{>{\hangindent=2em}Xrrrr}
    \begin{tabular}{l@{}rrrr}
    \toprule
        Topic &
        \begin{tabular}{@{}c@{}}no.\ flaky\\projects\end{tabular} &
        \begin{tabular}{@{}c@{}}no.\\projects\end{tabular} &
        \begin{tabular}{@{}c@{}}flakiness\\rate\end{tabular} &
        \begin{tabular}{@{}c@{}}average\\no.\ tests\\/ project\end{tabular} \\
    \midrule
    Scientific/Engineering :: Artificial Intelligence & 17                            & 259                         & \SI{6.6}{\percent} & 35.3 \\
                              Scientific/Engineering  & 47                            & 796                         & \SI{5.9}{\percent} & 82.3 \\
           Scientific/Engineering :: Bio-Informatics  & 15                            & 261                         & \SI{5.7}{\percent} & 83.1 \\
                                           Utilities  & 65                            & \num{1289}                        & \SI{5.0}{\percent} & 30.0 \\
                     Software Development :: Testing  & 21                            & 435                         & \SI{4.8}{\percent} & 23.2 \\
 Software Development :: Libraries :: Python Modules  & 100                           & \num{2116}                        & \SI{4.7}{\percent} & 82.7 \\
                                Software Development  & 29                            & 623                         & \SI{4.7}{\percent} & 60.2 \\
UNSPECIFIED                                           & \numFlakyProjectsWithoutTopic & \numAllProjectsWithoutTopic & \SI{4.5}{\percent} & 36.1 \\
                 Software Development :: Build Tools  & 16                            & 439                         & \SI{3.6}{\percent} & 17.4 \\
                   Software Development :: Libraries  & 33                            & 910                         & \SI{3.6}{\percent} & 34.8 \\
                                Internet :: WWW/HTTP  & 15                            & 420                         & \SI{3.6}{\percent} & 23.6 \\
    \bottomrule
% \end{tabularx}
    \end{tabular}
}

\end{table}

\subsection{RQ2: Types of Flakiness in Python}%
\label{sec:evaluation_rq2_type_of_flakiness_in_python}

\begin{table}[t]
    \centering
    \caption{Root causes of the flakiness we observed}%
    \label{tab:categorization}
    \begin{tabular}{llr}
        \toprule
            Root Cause                                 & relative                                   & total \\
        \midrule
            Infrastructure                             & \InfrastructureFlakyTestsOverAllFlakyTests & \numInfrastructureFlakyTests \\
            \rule{0pt}{2ex} \\
            Test Order Dependency                      & \OdFlakyTestsOverAllFlakyTests             & \numODFlakyTests \\
                \phantom{AAA}victims                   &                                            & \numODVictimFlakyTests          \\
                \phantom{AAA}brittles                  &                                            & \numODBrittleFlakyTests         \\
                \phantom{AAA}could not be analyzed     &                                            & \numODNotAnalysedFlakyTests     \\
            \rule{0pt}{2ex} \\
            Non-order-dependent & \NodFlakyTestsOverAllFlakyTests            & \numNODFlakyTests \\
            \phantom{A}sample       &             & 100 \\
                \phantom{AAA}Network               &  & \numNODNetwork\\
                \phantom{AAA}Randomness            &  & \numNODRandom\\
                \phantom{AAA}IO                    &  & \numNODIO\\
                \phantom{AAA}Time                  &  & \numNODTime\\
                \phantom{AAA}Async Wait            &  & \numNODAsyncWait\\
                \phantom{AAA}Concurrency           &  & \numNODConcurrency\\
                \phantom{AAA}Resource Leak         &  & \numNODResourceLeak\\
                \phantom{AAA}Test Case Timeout     &  & \numNODTestCaseTimeout\\
                \phantom{AAA}Unordered Collections &  & \numNODUnorderedCollection\\
                \phantom{AAA}Too Restrictive Range &  & \numNODTooRestrictiveRange\\
                \phantom{AAA}Platform Dependency   &  & \numNODPlatformDependency\\
                \phantom{AAA}Test Suite Timeout    &  & \numNODTestSuiteTimeout\\
        \bottomrule
    \end{tabular}
\end{table}

\cref{tab:categorization} shows the number of flaky tests we found for each category of flakiness.
%
% -- Infrastructure flakiness
% Observation
We found \numInfrastructureFlakyTests tests to be instances of infrastructure flakiness.
%
% -- OD flaky tests
% Observation
Furthermore, we encountered \numODFlakyTests tests, which are flaky due to order dependencies.
Running the order-dependent tests in isolation, we found a majority of them to be victims~(\numODVictimFlakyTests) and a minority to be brittles~(\numODBrittleFlakyTests).
Roughly ten percent of all order-dependent tests could not be analyzed, mostly due to our test framework being unable to find the specified test.
Common obstacles for test re-identification are inheritance between test case classes, complex parametrizations, and test IDs differing from JUnit-XML IDs.

% -- NOD flaky tests
In the following, we discuss the most prominent categories of NOD flakiness (\numNODFlakyTests tests) we found in our sample together with representative examples.
%
% Network example
The most common category in our sample is network.
\cref{fig:NOD_example_geocoder} shows an example for network flakiness taken from project \enquote{geocoder}, a Python library that takes a location in form of an address and returns its geographic coordinates using online services.
The test failed because of an HTTP request exceeding the given timeout of \SI{10}{\second}.

\begin{figure}[t]
    \centering
% Chunk05, geocoder.
% location = 'Ottawa, Ontario'
\begin{lstlisting}[language=Python]
def test_komoot_multi_result():
    g = geocoder.komoot(location, maxRows=3, timeout=10)
>   assert g.ok
    assert len(g) == 3

AssertionError: assert False
  where False = <[ERROR - HTTPConnectionPool(host='photon.komoot.de', port=80): Read timed out. (read timeout=10)] Komoot - Geocode [empty]>.ok
\end{lstlisting}
\vspace{-1em}
\caption{Network timeout causing flaky failure}%
    \label{fig:NOD_example_geocoder}
\end{figure}

% Randomness
We also found many flaky tests due to randomness, with an example
shown in \cref{fig:NOD_example_gamble}. The test is part of project
\enquote{gamble}, a library that implements functionalities concerning
cards and dice.  The depicted test creates a deck of cards, shuffles
them, and checks if the top-most card has changed. With a low, but
existing chance, this test will fail as despite the cards have been
shuffled, the top-most card might still be the same.

\begin{figure}[t]
    \centering
% Chunk05, gamble.
\begin{lstlisting}[language=Python]
def test_deck_init() -> None:
"""test that we can create a deck of cards"""
    deck = Deck(shuffle=False)
    # omitted
    last_top = deck.top
    deck.shuffle(times=10)
>   assert last_top != deck.top

AssertionError:
    assert <Card:7> != <Card:7>
    where <Card:7> = <Deck[46]>.top
\end{lstlisting}
\vspace{-1em}
    \caption{Randomness causing flaky failure}%
    \label{fig:NOD_example_gamble}
\end{figure}

% Time
A less common, but existing cause of flakiness, is the wrong usage of the system time.
\cref{fig:NOD_example_cronjob} presents an example for that, taken from project \enquote{cronjob}, an API for the UNIX job scheduler cron.
The test creates a cronjob running every five minutes and checks if there is at least one second between the current system time and the next execution of the job.
It fails if the next execution happens within the next second.

\begin{figure}[t]
    \centering
% Chunk03, cronjob.
% cr.interval is computed depending on the current time.
\begin{lstlisting}[language=Python]
def test_cronrule():
    tests = [
        # omitted
        ('*/5 * * * *', operator.gt, 0),
        # omitted
    ]
    for condition, op, result in tests:
        cr = CronRule(condition)
>       assert op(cr.interval, result)

AssertionError: assert False
    where False = <built-in function gt>(0, 0)
\end{lstlisting}
%     \caption{Test, which is flaky due to system time.
%     }%
\vspace{-1em}
    \caption{
        System time causing flaky failure
         % \texttt{cr.interval} computes time in seconds between the next execution and the current time.
    }
    \label{fig:NOD_example_cronjob}
\end{figure}

% class CronRule:
%     # omitted
%     def interval(self) -> int:
%         return self.parser.get()
%
% class CronParser:
%     # omitted
%     def get(self):
%         now = datetime.now()
%         next_time = croniter(self.rule, now).get_next(datetime)
%         return (next_time - now).seconds

% Async wait

Of all tests in our sample, \SI{3}{\percent} were flaky because they did not properly wait for an asynchronous call to complete before making assertions on it.
One of them is depicted in \cref{fig:NOD_example_piripherals}, which is part of project \enquote{piripherals}, a tool to interact with peripherals for the RaspberryPi.
As it uses a mocking framework, its tests can also be executed successfully on other hardware.
The test fails in case a handler, which is a mocked object, has not been called within 0.01 seconds.

% chunk07, piripherals
\begin{figure}[t]
    \centering
\begin{lstlisting}[language=Python]
def test_mpr121_irq(GPIO, bus):
  # omitted
  for i in range(13):
    dev.write_word(0, 1 << i)
    irq()
    sleep(0.01)
>   handlers[i].assert_called_once_with(True, i)
    # omitted

AssertionError: Expected "mock" to be called once. Called 0 times.
\end{lstlisting}
\vspace{-1em}
    \caption{Asynchronous waiting causing flaky failure}%
    \label{fig:NOD_example_piripherals}
\end{figure}

% Test case timeout + hypothesis
%
Fuzzing tools can also produce flakiness as shown in \cref{fig:NOD_example_humansort}:
The depicted test belongs to \enquote{humansort}, a tool that sorts filenames in a more human readable way.
The project uses \toolname{hypothesis}~\cite{MH19}, a property-based testing library for Python, which dynamically parametrizes a test case, searching for edge cases that might cause it to fail.
However, \toolname{hypothesis} itself can also cause test failures, like in this case, where a deadline was exceeded.
%
% hypothesis causes flakiness due to randomness
We also found cases in which \toolname{hypothesis} caused flakiness falling into the randomness category:
As its search for test input values is non-deterministic, it might find a bug in one execution, but not in the next one.
This can be avoided by using its builtin cache mechanism.

\begin{figure}[t]
    \centering
% chunk05, humansort
\begin{lstlisting}[language=Python,caption={}]
from hypothesis import given
from hypothesis.strategies import from_regex, integers, lists, tuples

strat_strings = from_regex(r"\A[^0-9]*\Z")
strat_mod = tuples(integers(), lists(integers(min_value=0), max_size=10))
strat = strat_strings, lists(strat_mod, max_size=5)

@given(lists(tuples(*strat)))
> def test_sort_property(e):
    # omitted

hypothesis.errors.DeadlineExceeded: Test took 201.15ms, which exceeds the deadline of 200.00ms
\end{lstlisting}
\vspace{-1em}
    \caption{Fuzzing framework causing flaky timeout}%
    \label{fig:NOD_example_humansort}
\end{figure}

% Multiple categories: IO, Unordered collection, Too restrictive range
%
Some cases also match multiple categories, like the one shown in \cref{fig:NOD_example_arlib}, which belongs to project \enquote{arlib}, a common interface for archive manipulation.
The test is flaky because the order in which file names are returned by \texttt{ar.member\_names}, which internally calls \texttt{os.listdir}, is not deterministic, however, it is compared against an ordered data structure.
The category of flakiness in which this test case falls, is debatable:
We classified it as flaky due to unordered collection, but it could also be counted as flaky due to IO, as the non-determinism involves the file system.
Furthermore, the flakiness could be removed by not insisting on a certain ordering between the elements, hence indicating a too restrictive range.
The case exemplifies, that the categories of flakiness are not distinct and one flaky test might match several categories.
\begin{figure}[b]
    \centering
% chunk02, arlib
\begin{lstlisting}[language=Python,caption={}]
def test_arlib_read(fname):
  if sys.version_info[0] >= 3:
    with arlib.open(fname, 'r') as ar:
>     assert ar.member_names == ['a.txt', 'b.txt']

AssertionError:
  assert ["b.txt", "a.txt"] == ["a.txt", "b.txt"]
\end{lstlisting}
\vspace{-1em}
\caption{Flakiness matching multiple categories (Unordered collection, IO, Too restrictive range)}%
    \label{fig:NOD_example_arlib}
\end{figure}

\summary{RQ2: What types of flakiness are prevalent in Python?}{\OdFlakyTestsOverAllFlakyTests of all flakiness we observed was caused by order dependencies, \InfrastructureFlakyTestsOverAllFlakyTests by infrastructure flakiness, and \NodFlakyTestsOverAllFlakyTests mostly and to equal degrees by the use of networking and randomness APIs.}

\begin{figure}[htpb]
    \centering
\begin{subfigure}[b]{\linewidth}
    \centering
    \includegraphics[width=\linewidth]{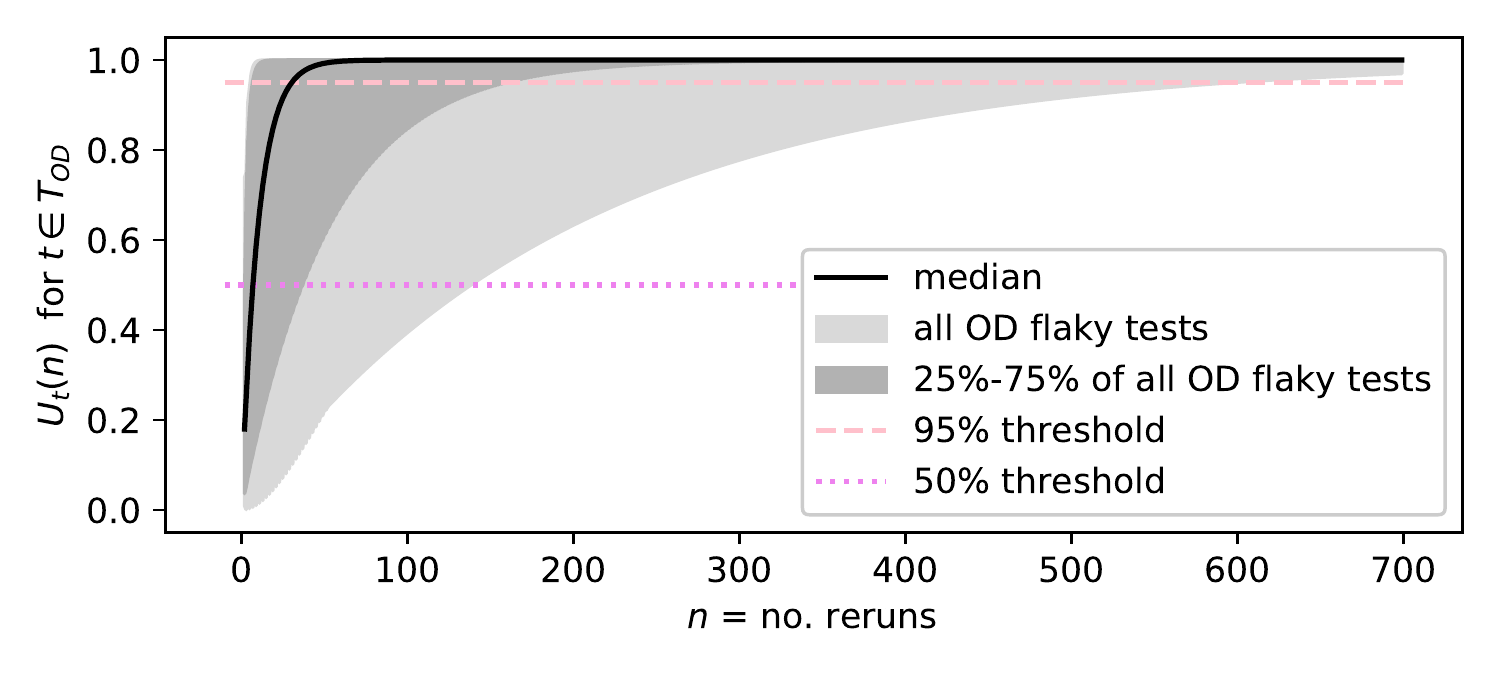}
    \caption{Chance to unveil flakiness after $n$ reruns for all OD flaky tests}%
    \label{fig:u_t_n_OD}
\end{subfigure}

\begin{subfigure}[b]{\linewidth}
    \centering
    \includegraphics[width=\linewidth]{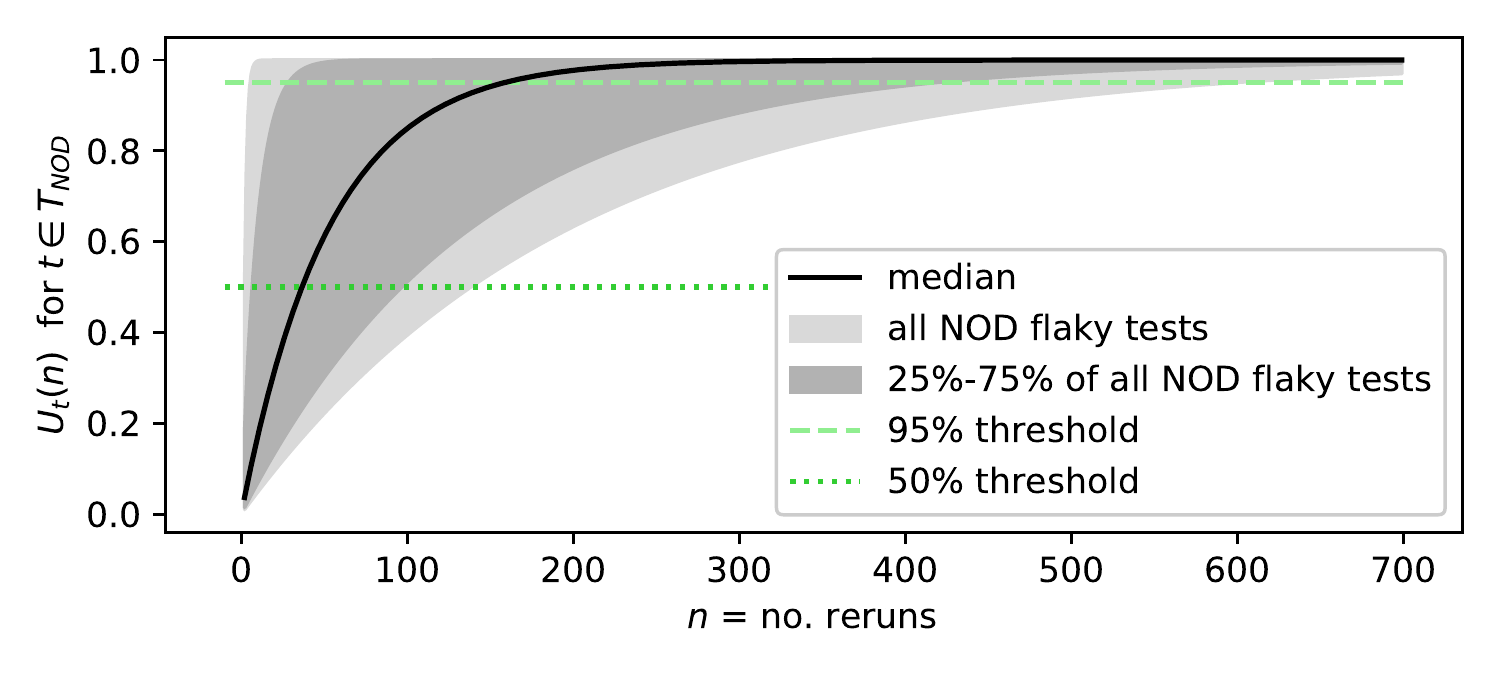}
    \caption{Chance to unveil flakiness after $n$ reruns for all NOD flaky tests}%
    \label{fig:u_t_n_NOD}
\end{subfigure}
    \caption{Statistical chance to unveil flakiness}%
    \label{fig:u_t_n}
\end{figure}

\subsection{RQ3: Degree of Flakiness}%
\label{sec:evaluation_rq3_degree_of_flakiness}

\cref{fig:u_t_n_OD} and \cref{fig:u_t_n_NOD} depict the cumulative
probability distribution of $U_t(n)$ of all OD tests
$t \in T_{OD}$ and NOD tests $t \in T_{NOD}$.
The average number of reruns needed to expose a test's
flakiness---which is the answer to question (a)---can now be derived
visually, as the intersection between the probability of exposing the
flakiness of an average flaky test (black line), and the
\SI{95}{\percent} threshold (dashed line).

\begin{figure}[tpb]
    \centering
\begin{subfigure}[b]{\linewidth}
    \centering
    \includegraphics[width=\linewidth]{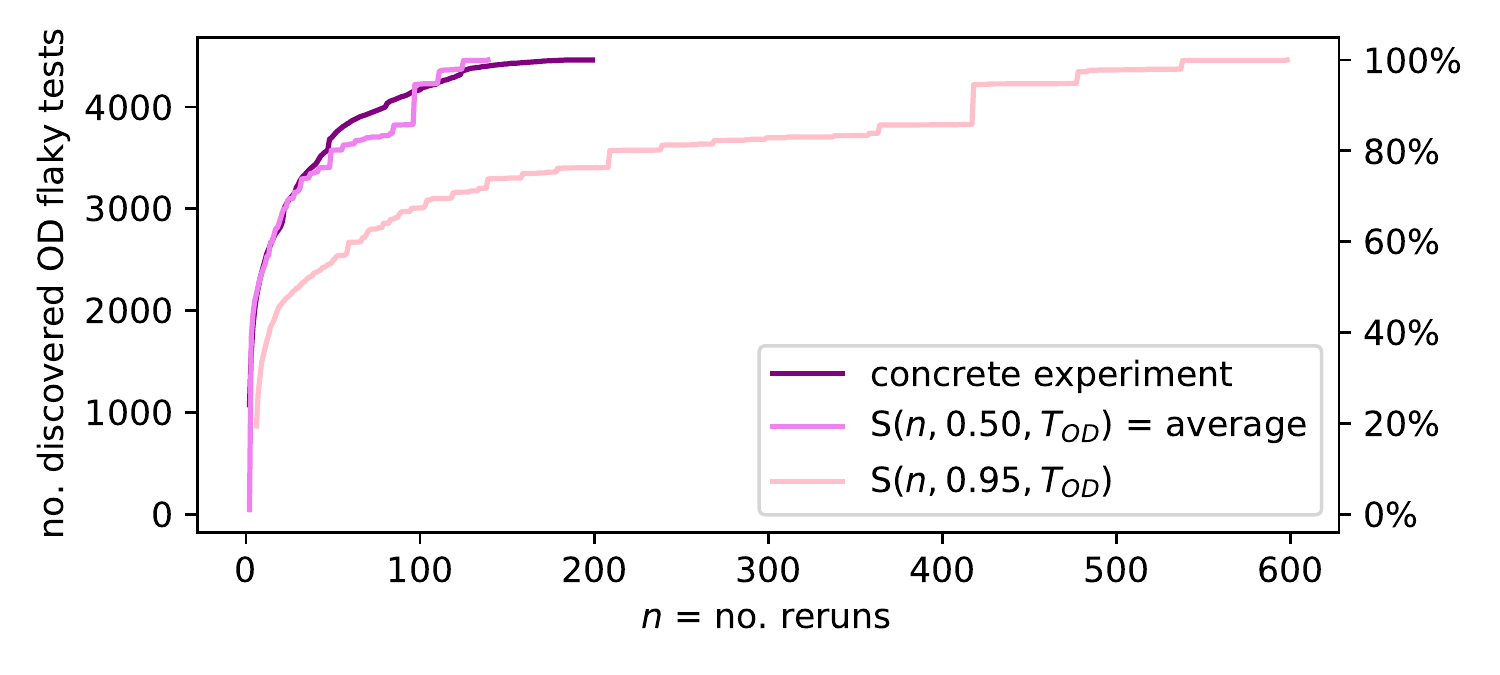}
    \caption{No.\ reruns necessary to unveil order-dependent flakiness}%
    \label{fig:reruns_OD}
\end{subfigure}

\begin{subfigure}[b]{\linewidth}
    \centering
    \includegraphics[width=\linewidth]{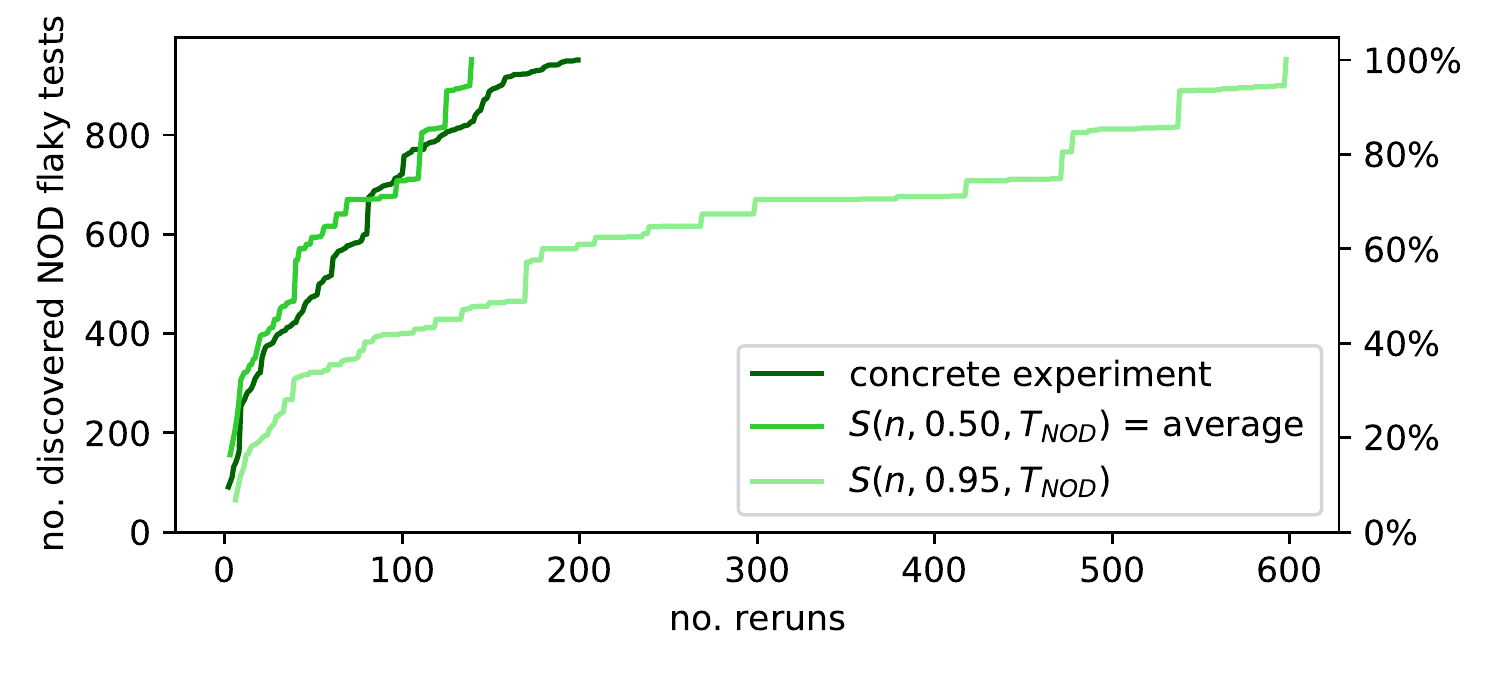}
    \caption{No.\ reruns necessary to unveil non-order-dependent flakiness}%
    \label{fig:reruns_NOD}
\end{subfigure}
\caption{Flakiness found after $n$ reruns}%
\label{fig:reruns}

\end{figure}

\cref{fig:reruns_OD} and~\cref{fig:reruns_NOD} show the sum of
flaky tests $S(n,x,T)$ found after $n$ reruns for all OD flaky tests
($T = T_{\mathrm{OD}}$) and all NOD flaky tests
($T = T_{\mathrm{NOD}}$). Metric $x = \mathrm{once}$ shows the number
of flaky tests we found in our concrete experiment, $x = 0.5$ is the
average number of flaky tests one can expect to find, and $x = 0.95$
is a more conservative estimation.

% Let's look at \cref{fig:reruns_OD}:
% % once after 100
% After \num{100} reruns, we discovered \num{3360} order-dependent flaky tests, which is indicated by the darkest of the three lines.
% %
% This insight, however, is based on a one-time experiment and will therefore be hard to reproduce.
% %
% The other two lines show the number of flaky tests one can expect to find with a certain confidence after a certain number of reruns, based on the pass- and failure-rates of all test executions we conducted.
% %
% % 50% and 95% after 100 runs
% With a confidence rate of \SI{50}{\percent} (lightest curve), the metric suggests, that after \num{100} reruns, one can expect to find \num{3389} flaky tests, which is almost the same number as we actually found.
% With a certainty of \SI{95}{\percent}, however, the metric suggests that one would only find \num{2703} flaky tests.
% %

% confirm: n_once ~ n_50%
For order-dependent flakiness, the number of flaky tests discovered in our concrete experiment is very close to the average-curve.
For NOD flaky tests, this relationship is weaker, however still strong, showing that the confidence gained from studying the number of tests needed to unveil flakiness once is only \SI{50}{\percent}.
%
% OD converges, NOD does not
\cref{fig:reruns_OD} shows that the majority of order-dependent flaky tests is discovered within the first 50 reruns, followed by a more steady growth, and finally almost saturation when exceeding 150 reruns.
In contrast to that, the rate at which new non-order-dependent flaky tests were discovered decreased only slightly even after 150 reruns.

With the help of \cref{fig:reruns} we are now able to answer all the researcher's questions introduced in Section~\ref{sec:study_setup_rq3_degree_of_flakiness}.
% Answers to previous questions
\begin{enumerate}[label=(\alph*)]
    \item In order to be \SI{95}{\percent} sure that a test case is not flaky due to non-order-dependent reasons, one would have to run it at least \numRerunsOneNODTestNinetyFive times.
    For order-dependent flakiness, this number is lower, at \numRerunsOneODTestNinetyFive random-order executions.
    % For the majority of NOD flaky tests, at least \numRerunsOneNODTestNinetyFive reruns were required to expose their flakiness with \SI{95}{\percent} confidence.
    %
    \item If you rerun your tests ten times, you should not expect to find more than \ratioNodFlakyTestsFoundAfterTenReruns of all NOD flaky tests and \ratioODFlakyTestsFoundAfterTenReruns of all OD flaky tests on average.
    \item Finding \SI{80}{\percent} of all NOD flakiness requires on average at least \numRerunsEightyPercentNodFlakyTestsFivety reruns; or at least \numRerunsEightyPercentNodFlakyTestsNinetyFive reruns, in case you want to be \SI{95}{\percent} sure.
    For OD flaky tests, these numbers are less than half as large with \numRerunsEightyPercentODFlakyTestsFivety reruns for \SI{50}{\percent} confidence and \numRerunsEightyPercentODFlakyTestsNinetyFive reruns for \SI{95}{\percent} confidence.
    %
%    \item To check, if a failure occurred due to flakiness, for the majority of NOD flaky tests, only \numRerunsUnveilNODFlakyFailureNinetyFive rerun is needed to achieve a statistical confidence of \SI{95}{\percent}.
%    For OD flakiness, \numRerunsUnveilODFlakyFailureNinetyFive reruns are needed to achieve the same confidence.
%    If the test keeps failing after several reruns, it might still be flaky due to infrastructure flakiness, which often manifests itself in failure bulks.
\end{enumerate}

Answer (c) suggests that even after \num{200} reruns, we did not find all flaky tests in our dataset.
To estimate the percentage of flakiness we found, we calculate the fraction of flaky tests one can expect to find after 200 reruns with a \SI{95}{\percent} confidence:
\begin{align*}
    \frac{S(200, 0.95, T_{\mathrm{NOD}})}{|\ T_{\mathrm{NOD}}\ |} = \frac{580}{952} \approx \SI{61}{\percent}
\end{align*}
Therefore, with a \SI{95}{\percent} chance, we found at least \SI{61}{\percent} of all NOD flakiness in our dataset.
This statement is based on the assumption, that the flaky tests still hidden in the test suite behave similarly to the flaky tests we were able to expose, with regard to their passing and not-passing rates.

The practitioner's question is answered based on the passing probability $P_t(\mathrm{pass})$ for $t \in T_{\mathrm{NOD}}$ and $t \in T_{\mathrm{OD}}$:
  \begin{enumerate}[label=(\alph*)]
    \item[(d)] To check if a failure occurred due to flakiness, for the majority of NOD flaky tests, only \numRerunsUnveilNODFlakyFailureNinetyFive rerun is needed to achieve a statistical confidence of \SI{95}{\percent}.
    For OD flakiness, \numRerunsUnveilODFlakyFailureNinetyFive reruns are needed to achieve the same confidence.
\end{enumerate}
This means that common practice of failure reruns~\cite{Micco2016} is
generally sufficient
(when ignoring infrastructure flakiness, which often manifests itself in failure bulks).

% We do not give answers for questions (a) to (c) for order-dependent flaky tests, because
% (1) the number of reruns required to expose the flakiness substantially depends on the size of the test suite.
% Reporting the same rerun-statistics as we did for NOD flaky tests would therefore indirectly report the number of other tests in the test suite.
% (2)
% for OD flaky tests, smart identification strategies exist to minimize the number of reruns needed~\cite{GBZ18}.
%
% However, we can still conclude that, on average, exposing order-dependent flakiness requires fewer reruns than exposing non-order dependent flakiness.
%

% Conclusion
While recommending an ideal number of reruns remains an inherently complex problem, our results still show that finding flakiness via rerun requires a tremendous amount of test executions, which should be a further motivation to find other ways of exposing flakiness.

\summary{RQ3: What is the degree of flakiness of flaky tests in Python?}{Flaky tests in Python have a low failure rate, resulting in a low number of reruns necessary to check, if a failure was flaky (\numRerunsUnveilNODFlakyFailureNinetyFive rerun for \SI{95}{\percent} confidence on NOD flaky tests), but a high number of reruns necessary to check if a test in general contains flakiness (\numRerunsOneNODTestNinetyFive reruns for \SI{95}{\percent} confidence on NOD flaky tests).}

% Figure~\ref{fig:flaky_tests_od_nod} shows the number of OD and NOD flaky tests discovered by our investigation and by the authors of iDFlakies.
% \begin{figure}[tpb]
%     \centering
%     \includegraphics[width=\linewidth]{flaky_tests_od_nod}
%     \caption{Flaky Tests categorized by order-dependent and not-order-dependent}%
%     \label{fig:flaky_tests_od_nod}
% \end{figure}

% TODO How flaky are the tests
% - how is flakiness distributed amongst projects -> what is the variance? (assumption: high variance, either no test is flaky at all or there are multiple flaky tests)

% TODO export charts as pgf -> set width ~=3.5 = 1/2* textwidth
% textwidth in inch: \printinunitsof{in}\prntlen{\textwidth}

%%% Local Variables:
%%% mode: latex
%%% TeX-master: "../main"
%%% End:

%% file: sections/50_related.tex
\section{Related Work}\label{sec:related}

% We chose this study, as it features the largest number of projects amongst the prior work we considered:

The largest study on test flakiness in Java is the
\toolname{iDFlakies}~\cite{LOS+19} study.
In this study, a set of 683
Java projects, consisting of 639 popular GitHub projects and 44
projects from previous studies, was screened for flakiness.
While a total of \num{1974084} tests were reported for the overall dataset, only \num{945} out of \num{5171} modules were analyzed.
To allow for a better comparison, we determined that this dataset contains \num{89568} test cases.
\cref{tab:comparison_idflakies} puts these and our findings in contrast.
%
% higher rate of flaky projects as expected
While the rate of projects containing at least one flaky test is
higher in the Java dataset, this can be attributed to the biased
sampling technique by which the Java projects were chosen, as all
\num{26} projects taken from Bell et al.~\cite{BLH+18} are chosen specifically
because they contain flakiness.
%
% similar rate of flaky tests
Looking at the total number of flaky tests discovered, on the other
hand, both datasets exhibit a similar rate of around 1 in every
\num{120} (this study) vs. 1 in every \num{210} (iDFlakies) test being
flaky.
%
% The rates are even more similar, when leaving out infrastructure flakiness, which is not considered by the authors of~\cite{LOS+19}.
% % Infrastructure
% Infrastructure flakiness would also manifest itself earlier in Java, as a Java project would not even compile in case the dependencies were not correctly installed.
%
% Root causes
We also see that order-dependency seems to be a more pressing issue in
Python while non-order-dependency is rarer.
% Reviewer critic: "If the goal of the study was to compare flakiness in Java and in Python, it would have been important to use the exact same setup as iDFlakies."
Note, however, that our study does not use the exact same setup as the iDFlakies study, as we conduct more reruns without using smart scheduling.

\begin{table}[t]
    \caption{Comparison: iDFlakies dataset \& our dataset}
    \label{tab:comparison_idflakies}
    \centering
    \resizebox{\columnwidth}{!}{
    \begin{tabular}{lrrrrrrr}
    \toprule
    Study        & Projects & Tests & \multicolumn{5}{c}{Flaky} \\
                 & &
                 & Projects
                 & Tests
                 & OD
                 & NOD
                 & Infr. \\
    \midrule
    iDFlakies
        & \num{683}
        & \num{89568}
        & \SI{12}{\percent}
        & \SI{0.47}{\percent}
        & \SI{50}{\percent}
        & \SI{50}{\percent}
        & -     \\
    This study
        & \numAllProjects
        & \numAllTests
        & \FlakyProjectsOverAllProjects
        & \FlakyTestsOverAllTests
        & \OdFlakyTestsOverAllFlakyTests
        & \NodFlakyTestsOverAllFlakyTests
        & \InfrastructureFlakyTestsOverAllFlakyTests     \\
    \bottomrule
    \end{tabular}}
\end{table}

% NOD root causes
A fine-grained categorization of non-order-dependent flaky tests was
previously performed on \num{400} known flaky tests found in the
commit history of Apache projects~\cite{LHE+14} and the Mozilla issue
tracker~\cite{EPC+19}.
\cref{tab:categorization_comparison} demonstrates that, whereas they
reported Async Wait and Concurrency to be the most common root causes of
NOD flakiness, we found only little evidence of
these categories. On the other hand, the categories causing most NOD
flakiness in our dataset (namely Networking and Randomness),
played only a minor role in the two other studies.

\begin{table}[t]
    \centering
    \caption{Root causes of non-order-dependent flaky tests
        % we observed in comparison to similar prior studies
    }%
    \label{tab:categorization_comparison}
    \begin{tabular}{lrrr}
        \toprule
        Root Cause                & This study                   & ~\cite{LHE+14} & ~\cite{EPC+19} \\
        \midrule
        Async Wait                & \numNODAsyncWait             & 74 & 52\\
        Concurrency               & \numNODConcurrency           & 32 & 61 \\
        Resource Leak             & \numNODResourceLeak          & 11 & 14\\
        Network                   & \numNODNetwork               & 10 & 0 \\
        Time                      & \numNODTime                  & 5  & 4\\
        IO                        & \numNODIO                    & 4  & 0\\
        Randomness                & \numNODRandom                & 4  & 3\\
        Floating Point Operations & -                            & 3  & 6 \\
        Unordered Collections     & \numNODUnorderedCollection   & 1  & 0\\
        Too Restrictive Range     & \numNODTooRestrictiveRange   & -  & 40\\
        Test Case Timeout         & \numNODTestCaseTimeout       & -  & 18\\
        Platform Dependency       & \numNODPlatformDependency    & -  & 10\\
        Test Suite Timeout        & \numNODTestSuiteTimeout      & -  & 4\\
        Infrastructure            & \numInfrastructureFlakyTests & -  & - \\
        \bottomrule
    \end{tabular}
\end{table}

% iFixFlakies
Order-dependent flakiness was investigated in detail by Shi et
al.~\cite{SLO+19}, who found \num{100} victims and \num{10} brittles
in a set of \num{110} order-dependent Java tests.  The same trend
towards far more victims than brittles can also be observed in our
results.
One reason why brittles are rarer than victims might be that in
modern test frameworks state-setters can be declared actively, which
restricts possible test orders and therefore avoids brittles.

% % Possible explanation
% Attempting to give an explanation for the observable trend, one might argue that it is easier to, by mistake, create a victim than it is to create a brittle.
% A brittle requires a separate test case serving as a state-setter that must run before the brittle in order for it to pass.
% As using state-setters is a common practice, test frameworks offer special ways to declare such fixtures.
% Using such framework functionality, however, does not create brittles as the framework schedules the test execution in a way that the state-setter will always be executed before the test case, even when running the tests in random order.
% The creation of brittles therefore indicates that the developer felt the need to use state-setters but was unaware of existing solutions or used the existing solution in a wrong, unintended way.
% Creating victims on the other hand is easy, as it just requires one test to write to the global state (polluter) and another test to read and assert on the same global state (victim).

% Zhang et al.~\cite{ZJW+14} studied 96 order-dependent tests found in 5 issue trackers.

% Overall, our results align with the one's of previous work, showing that flakiness is equally present in Python projects as in Java project, despite the root causes being different, especially within non-order-dependent flakiness.

% on RQ3
Zhang et al.~\cite{Zhang2006} proposed a technique to predict the
failure rate of software before deployment. Although we also used
probabilistic models, our aim is to derive sound recommendations on
how to determine flakiness. Our methodology to classify flaky tests
involved traces, which were also used to localize error
causes~\cite{Ball2003}. Test flakiness may have knock-on effects on
other aspects such as test prioritization~\cite{ZJW+14}, mutation
analysis~\cite{SBM19}, or build crashes~\cite{RR18}, and for research
purposes flakiness can also be seeded~\cite{CRP+19}; since our paper represents the first study on flakiness in Python, we focused on a basic understanding of flakiness before considering such applications.

%\toolname{FlakiMe}~\cite{CRP+19} simulates the effects of non-deterministic test behavior on other techniques under laboratory controlled conditions, by seeding flakiness into a given test suite.

%%% Local Variables:
%%% mode: latex
%%% TeX-master: "../main"
%%% End:

%% file: sections/60_conclusions.tex
\section{Conclusions}\label{sec:conclusions}

Flaky tests represent a fundamental challenge in modern software
development. While previous research investigated this problem
predominantly in the context of Java software development,
we demonstrated that flaky tests are equally
prevalent in the Python ecosystem. Using a dataset of \numAllProjects projects
from the Python package index we found \numFlakyProjects projects in
which flakiness exists and a total of \numFlakyTests flaky tests. Although this
number is comparable to prior findings on Java, the reasons for this
flakiness differ, which is indicative of the different target domains
for Python, such as scientific software or web applications.

Besides a demand to extend existing techniques and tooling also to the
Python environment, these findings suggest that future work on the
peculiarities on flakiness in Python will be required, for example to
address flakiness in machine learning or scientific software. To this
purpose, we provide our dataset~\cite{gruber_martin_2021_4450435} and the
tooling\footnote{
    \url{https://github.com/se2p/FlaPy}, accessed 2021--01--18
} as open source to the community, hoping to foster research on new techniques to
automatically identify, classify and eliminate flakiness.
We encourage the investigation of flakiness in other upcoming languages such as Go or Rust as well as a more detailed exploration of infrastructure flakiness.

%%% Local Variables:
%%% mode: latex
%%% TeX-master: "../main"
%%% End: